\documentclass[12pt,onecolumn,draftclsnofoot]{IEEEtran}

\usepackage{times}
\usepackage[final]{graphicx}
\usepackage{amsmath,amsfonts}
\usepackage{amssymb,amsbsy}
\usepackage{times}

\usepackage{epsfig}
\usepackage{cite}
\usepackage{stfloats}
\usepackage{subfigure}

\newcommand{\hspp}{\hspace{0.05in} }
\newcommand{\hsppp}{\hspace{0.02in} }

\newcommand\ignore[1]{}

\newtheorem{thm}{Theorem}

\title{
Mitigating Hand Blockage with Non-Directional Beamforming Codebooks 
\author{Vasanthan Raghavan, Ricardo A. Motos, M. Ali Tassoudji, Yu-Chin Ou, \\
Ozge H.\ Koymen, and Junyi Li \\
Qualcomm Technologies, Inc., USA
}
\thanks{A shorter version of this paper was presented at the IEEE International 
Conference on Communications (ICC), Montreal, Canada, June 2021~\cite{vasanth_icc21}.} }

\begin{document}
\maketitle

\begin{abstract} 
\noindent 
Hand blockage leads to significant performance impairments at millimeter wave carrier frequencies. A number of prior works have characterized the loss in signal strength with the hand using studies with horn antennas and form-factor user equipments (UEs). However, the impact of the hand on the effective phase response seen by the antenna elements has not been studied so far. Towards this goal, we consider a measurement framework that uses a hand phantom holding the UE in relaxed positions reflective of talk mode, watching videos, and playing games. We first study the impact of blockage on a directional beam steering codebook. The tight phase relationship across antenna elements needed to steer beams leads to a significant performance degradation as the hand surface can distort the observed amplitudes and phases across the antenna elements, which cannot be matched by this codebook. To overcome this loss, we propose a non-directional beamforming codebook made of amplitudes and/or quantized phases with both these quantities estimated as necessary. Theoretical as well as numerical studies show that the proposed codebook can de-randomize the phase distortions induced by the hand and coherently combine the energy across antenna elements and thus help in mitigating hand blockage losses.
\end{abstract}

\section{Introduction}
Millimeter wave systems have significantly matured over the last five years with advances in
technological aspects, low-complexity and low-cost manufacturing as well as in standard
specifications and regulatory support. Enabled by these advances, the first wave of commercial
deployments in the $28$ and $39$ GHz regimes are now currently available in the market across multiple
geographies. Yet, a number of basic issues in terms of practical viability of systems operating at
millimeter wave frequencies are still not very well understood. One such issue is the question
of hand blockage that can significantly impair link margins at millimeter wave frequencies.

Modeling of blockage has received significant attention over the last few years. For example,
ray-tracing based blockage models have been proposed for 802.11(ad)~\cite{802d11_maltsev} as 
well as for the Third Generation Partnership Project (3GPP) Fifth Generation-New Radio (5G-NR)
systems~\cite{3gpp_CM_rel14_38901}. In particular, the 3GPP 5G-NR model captures
the spatial region of blockage in a local coordinate system around the user equipment (UE) in the
Portrait and Landscape modes with a $30$ dB flat loss assumed over this region. It is now
understood that the $30$ dB loss is {\em pessimistic} and is mostly a reflection of horn antenna 
studies such as~\cite{maccartney_2017,maccartney_2017_gcom} that were used to initiate discussions on 
the blockage model at 3GPP. More recent studies that use 
phased array systems show a considerably smaller blockage loss than $30$ dB.

For example, blockage studies at $15$ GHz with subarray/antenna module diversity that allows module switching
across different paths/clusters in the channel is shown to result in reduced blockage losses~\cite{zhao_ericsson,zhao_ericsson2}. 
A similar study with ray tracing is performed in~\cite{zhao_ericsson3} for $15$ and
$28$ GHz systems. 
Creeping waves and diffraction of signals are attributed as reasons for reduced blockage losses
at $28$ GHz in~\cite{syrytsin}. Simulated studies of hand blockage losses with an $8 \times 1$ 
linear antenna array and a $10 \times 1$ irregular antenna array in a $28$ GHz form-factor phone 
design are presented in~\cite{yu} and~\cite{xu}, respectively. In both works, reduced blockage losses
relative to the 3GPP model are reported. Simulation studies of the finger at $60$ GHz are reported
in~\cite{haneda1,haneda3} and a large loss variation is reported depending on the finger
placement on/near the antenna module. User effects on the power variation are reported for
$21.5$ GHz systems in~\cite{hejselbaek} with many scenarios of loss and some scenarios of gain
observed. Phased array vs.\ switched diversity array tradeoffs with hand and body blockage are studied 
in~\cite{syrytsin1} and the regime where each approach is better is quantified. 

In some of our prior 
works~\cite{vasanth_blockage_tap2018,vasanth_comm_mag_18}, a $28$ GHz form-factor prototype with 
3GPP-type beam management solution was used to study the impact of blockage losses. A median blockage 
loss of no more than $\approx 15$ dB was reported even with the hardest hand grip and this reduced 
loss was attributed to the increased beamwidth of the beam ($\approx 25^{\sf o}$ for a $4 \times 1$ 
array) relative to a horn antenna setup ($\approx 8^{\sf o}$ to $10^{\sf o}$). The increased beamwidth 
allows more energy to be collected by the phased array even with the presence of a hand thereby reducing 
the effective blockage losses. An important caveat common to these studies is that they are either based 
on ray-tracing or electromagnetic simulation studies, or with experimental prototypes that may/may not 
be a form-factor implementation.
 
More recently, measurement based blockage estimates with a commercial grade $28$ GHz UE design are reported
in~\cite{vasanth_blockage_2020}. Here, loss of less than $10$ dB and $20$ dB are estimated for loose
and hard hand grips, respectively. Further, it is shown that blockage can lead to reflection-associated
gains in certain directions over the sphere and these scenarios are important for loose hand grips.
The {\em region of interest} (RoI) where blockage loss/gain is relevant is identified and
spherical coverage improvement with blockage in loose hand grip mode is formally characterized. 

In general, if a blockage-driven link deterioration is seen, the UE can mitigate these
losses by either~\cite{vasanth_comm_mag_18}:
\begin{itemize}
\item Switching to a better path/cluster in the channel, which implicitly assumes multiple
capabilities at the UE side, or
\item Living with the deteriorated path and the concomitant link degradation.
\end{itemize}
For the first approach (beam switching) to work, we implicitly rely on a potentially densified network with viable paths/clusters
from multiple transmit nodes, a rich multipath channel with the connected (or potentially switching)
transmit node, multiple antenna modules and associated radio frequency integrated circuits (RFICs) to
allow the UE to switch to a different
path/cluster, beam switching latencies that are not disruptive to communications, and coordination
with the transmit node to allow beam switching~\cite{vasanth_comm_mag_18}. In this work, we propose an additional and entirely 
different mitigation strategy to add to this arsenal.

Towards this goal, we first report controlled blockage measurements with a commercial grade $28$ GHz
system using a $4 \times 1$ dual-polarized patch array with a commercial grade hand phantom in an 
anechoic chamber. In contrast to all the prior works that focus {\em only} on the amplitude response 
with blockage, we record the {\em complete} electric field (or array response) measurements that 
includes both amplitudes and phases in Freespace over the sphere. This set of benchmark measurements are
then compared with the complete 
hand blockage measurements using an anthropomorphic hand phantom in four
scenarios where the hand phantom is placed on top of the antenna module with one or two fingers
blocking/obstructing the antenna elements, and where the hand phantom is placed $1$ mm away from
the antenna array with signals blocked using one or two fingers. These scenarios capture
practical hand holdings such as those used in talk mode, watching videos, playing video games, etc.

We first consider an ideal/optimal maximum ratio combining (MRC) scheme that can tune the amplitudes and
phases across all the angles in the RoI with infinite precision. With this approach, we show that
median blockage losses in the range of $2.7$ to $4.5$ dB (larger losses corresponding to the two fingers of
the phantom hand on top of the antenna module) are seen with the $90$-th percentile losses being in the
range of $7.1$ to $11.6$ dB. Since infinite-precision codebooks are not practically viable, we then
consider a static beam steering codebook of size-$4$ (commensurate with the size of the antenna array).
We show that this approach incurs a median and $90$-th percentile losses in the range of $2.0$ to $2.3$ dB
and $3.7$ to $3.9$ dB, respectively. Bridging this gap in performance with any other approach is important
for the evolution of millimeter wave technology.

In this direction, we carefully study the amplitude and phase response with the hand phantom and
illustrate that multiple reflections from the different parts of the hand lead to a randomization
of the phase response as seen in the Freespace mode. As a result, a static codebook with a phase
response tailored to steer beams in specific directions may not lead to a constructive addition of
signals seen by the different antenna elements. This observation motivates a codebook enhancement
strategy that searches over different amplitude and phase shifter combinations so that a good
choice of beam weights can be arrived at. This good choice leads to an appropriate weighting of the
antenna elements to de-randomize the phases. Such an approach is shown to theoretically improve
the beamforming performance significantly. This result is also verified via measurements and
median and $90$-th percentile losses (relative to MRC) on the order of $\approx 0.35$ and $\approx 0.85$ dB, 
respectively, are reported. Note that these loss ranges are {\em far smaller} than those seen with a 
static codebook strategy.


This paper is organized as follows. Section~\ref{sec2} describes the measurement setup including the
experiments performed in this paper. Sections~\ref{sec3} and~\ref{sec4} study the impact of blockage
at a single antenna level and with beamforming, respectively. Section~\ref{sec5} proposes a mitigation
strategy to handle blockage and the intuitive motivation behind this specific choice. Theoretical
and numerical studies of the proposed approach are also considered with concluding remarks in
Section~\ref{sec6}.

\section{Measurement Setup}
\label{sec2}

\begin{figure*}[htb!]
\begin{center}
\begin{tabular}{ccc}
\includegraphics[height=1.8in,width=1.8in]{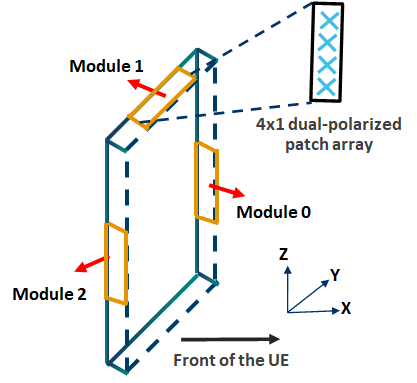}
&
\includegraphics[height=1.9in,width=1.7in]{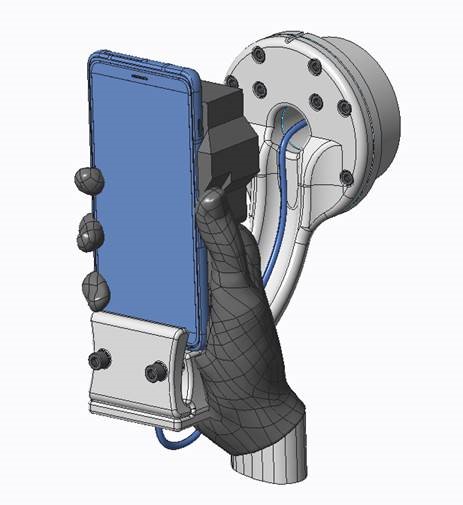}
&
\includegraphics[height=1.9in,width=3.0in]{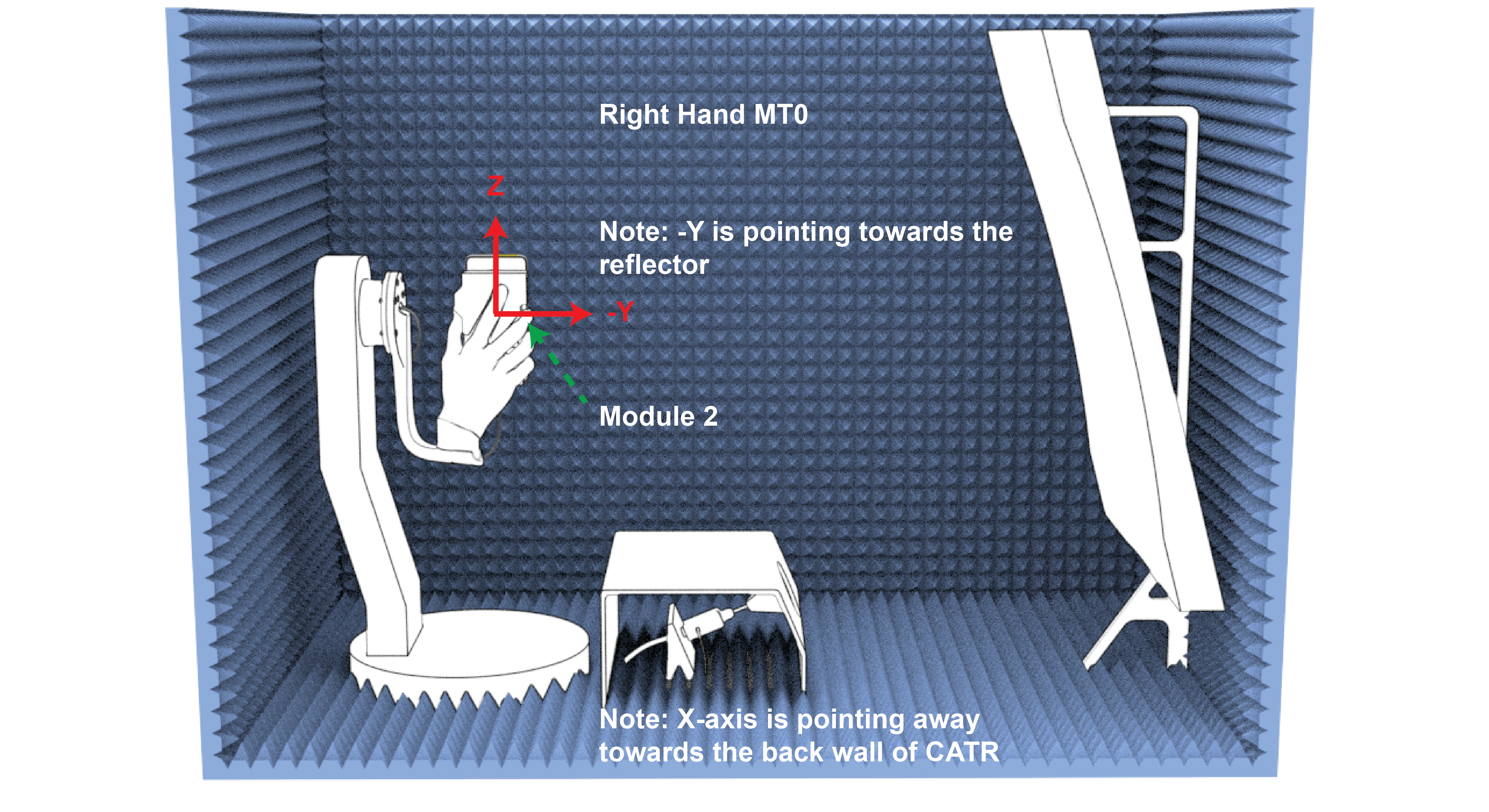}
\\
(a) & (b) & (c)
\end{tabular}
\caption{\label{fig_setup}
(a) Illustration of UE with antenna module structure. (b) Illustration of the hand
phantom holding the UE in (b) a stand-alone setup and (c) with the measurement setup in the
chamber. }
\end{center}
\end{figure*}

The UE considered in this study is equipped with a commercial grade millimeter wave modem
operating at $28$ GHz and using a 3GPP Rel.\ $15$ and $16$ standard specifications-compliant software
stack that performs intelligent beamforming and beam tracking. From an antenna module perspective,
as illustrated in Fig.~\ref{fig_setup}(a), the UE consists of three modules,
denoted as Modules $0$ to $2$, which are placed on the right long edge, top short edge and left long edge,
respectively, as seen from the front of the UE.
From a beamforming perspective, each 
antenna module has a $4 \times 1$ dual-polarized patch array that allows dual-polarized transmissions
via two radio frequency (RF) chains at $28$ GHz.

For the studies in this paper, we use a commercial grade anthropomorphic hand 
phantom~\cite{speag_hand_phantom} 
which is specifically designed for evaluating and optimizing over-the-air (OTA) performance of 
ultra-wide mobile phone devices (defined as having a width between $72$ and $92$ mm) such as the 
one considered in this paper. The hand phantom is manufactured using a
silicone-carbon-based mixture with material properties conforming to the Cellular Telecommunications
Industry Association (CTIA) definitions and standards for hand phantoms. The use of a special
low-loss silicone coating 
extends its useable frequency range from $3$ GHz to $110$ GHz. For the accuracy and reliability of the
evaluation results, as illustrated in Fig.~\ref{fig_setup}(b), the UE is placed on a specially designed
holding fixture that does not block the antenna modules. This fixture is made of a low-loss production
grade thermoplastic material and is 3D-printed using fused deposition modeling technology~\cite{fdm}.


While hand blockage studies are best performed with a true human holding the UE~\cite{vasanth_blockage_2020},  
avoiding the conflation of hand blockage effects with that of body blockage effects as well as 
exposure considerations suggest that studies with a hand phantom are a good proxy to capturing the true 
hand blockage effects. Therefore, to study the impact of hand blockage, four controlled studies 
corresponding to specific hand phantom positions are considered in this paper. These four positions include:
\begin{itemize}
\item Hand phantom on top of the antenna elements (that is, a $0$ mm air gap between the antenna module and 
the phantom) with either $1$ or $2$ fingers blocking/obstructing the antenna elements of the antenna module 
as illustrated in Fig.~\ref{fig_setup}, and
\item Hand phantom with a $1$ mm air gap from the antenna module with either $1$ or $2$ fingers
obstructing the radiation of the antenna module.
\end{itemize}
The $1$ mm air gap is introduced (and enforced) by placing a small rectangular low-loss foam material between 
the tip of the finger(s) and the antenna module. The $0$ mm air gap scenario is expected to capture a hand 
holding in talk mode as some of the fingers are expected to be touching the antenna module when a user talks 
on the phone. The $1$ mm air gap scenario is
expected to capture a user playing a game with a small air gap between the fingers and the UE. These
illustrative (but non-limiting) examples are thus expected to cover an interesting gamut of practical
applications. In the studies considered in this paper, the hand phantom is placed over Module 2 (antenna
module of interest) with a boresight direction of -Y axis (left long edge), as illustrated in
Fig.~\ref{fig_setup}(c).

For 3GPP-compliant OTA tests, the anechoic chamber uses a Compact Antenna Test Range
(CATR) method for far-field electromagnetic wave characterization. In this setup, a parabolic reflector is used to
collimate radiation at a test probe (as illustrated in Fig.~\ref{fig_setup}(c))~\cite{keysight_whitepaper}.
With this setup, automated chamber measurements (without the presence of humans) are conducted
to remove the across-human variations and the impact of body blockage in the collected data. Further, 
a full set of measurements for each antenna element can take a significant amount of time ($18$ to $20$ 
minutes) and automation of this process allows us to minimize human exposure to electromagnetic wave 
radiation. 

In this process, electric field information (amplitudes and phases) of each of the antenna elements of
the $4 \times 1$ array are separately collected in different modes of operation (Freespace mode as
well as with the hand phantoms correctly installed over the antenna module). Note that the electric
field information captures the array response as seen from the UE end including the effects of
housing, antenna substrate, metal/plastic, sensors, other electronic components, etc. The performance
of different analog beamforming codebook designs are then studied offline (as described in the sequel)
with the collected electric field information for all the antenna elements.

\section{Amplitude Response at Individual Antenna Level}
\label{sec3}




\begin{figure*}[htb!]
\begin{center}
\begin{tabular}{cccc}
\includegraphics[height=1.7in,width=1.5in]{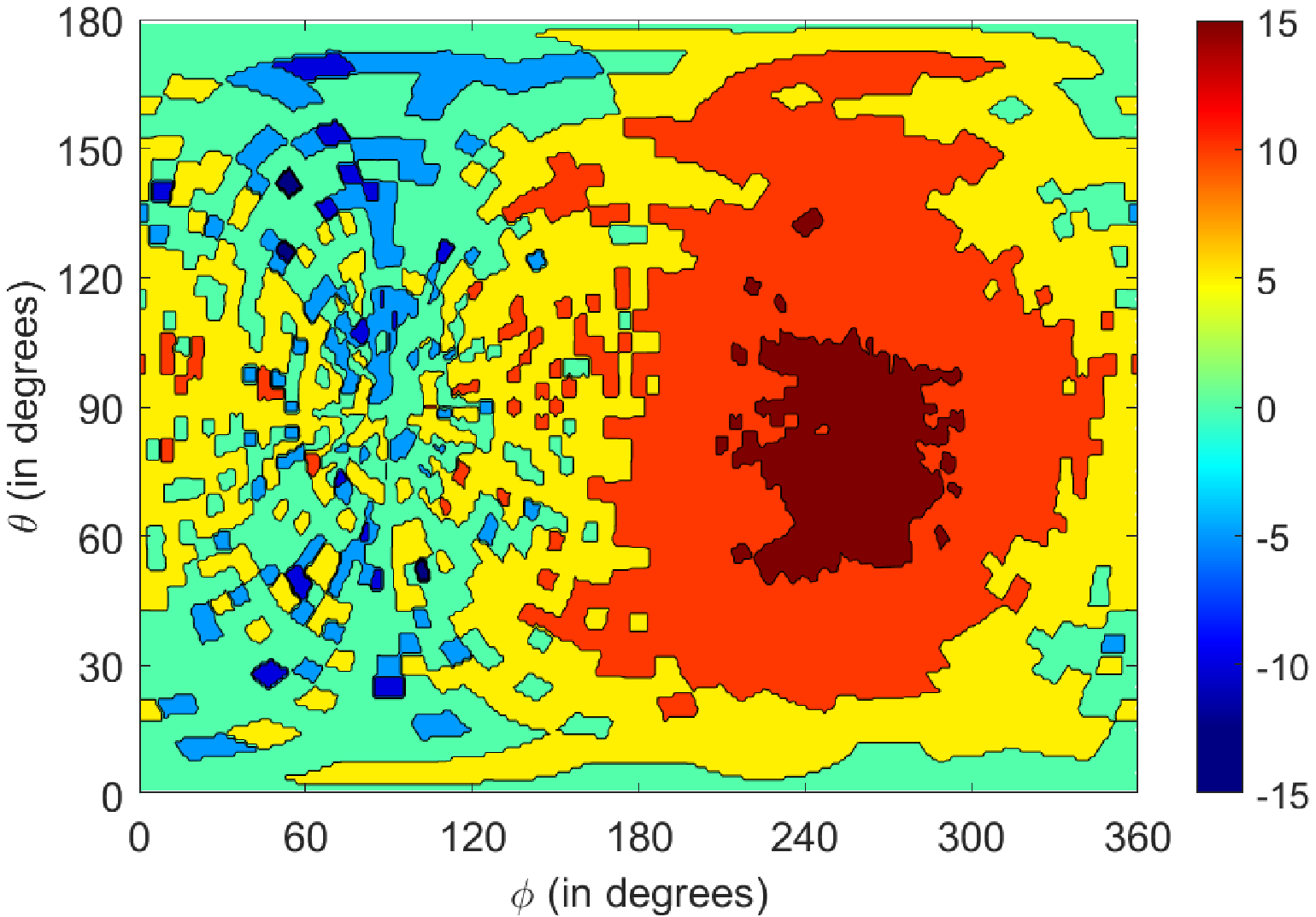}
&
\includegraphics[height=1.7in,width=1.6in]{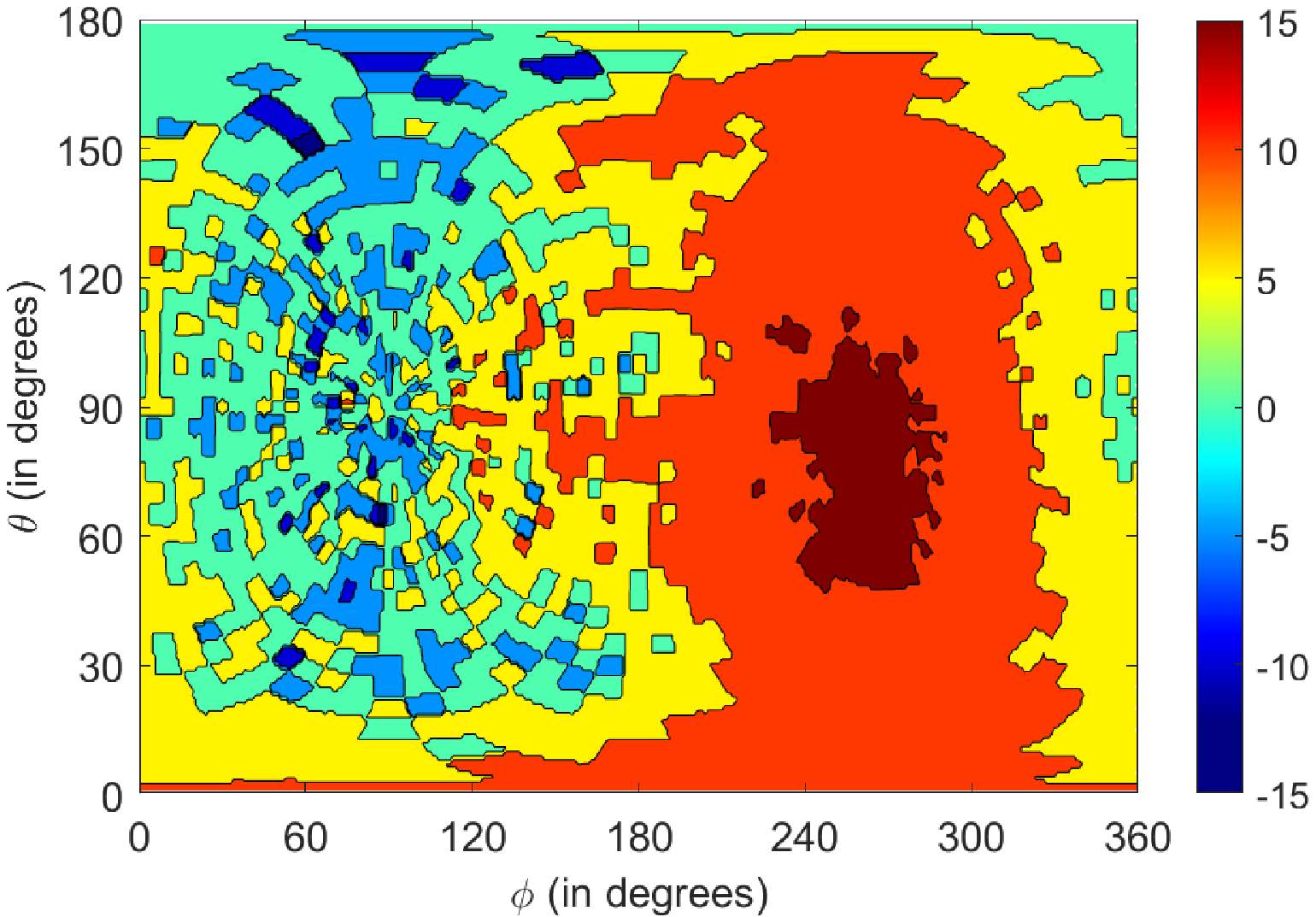}
&
\includegraphics[height=1.7in,width=1.6in]{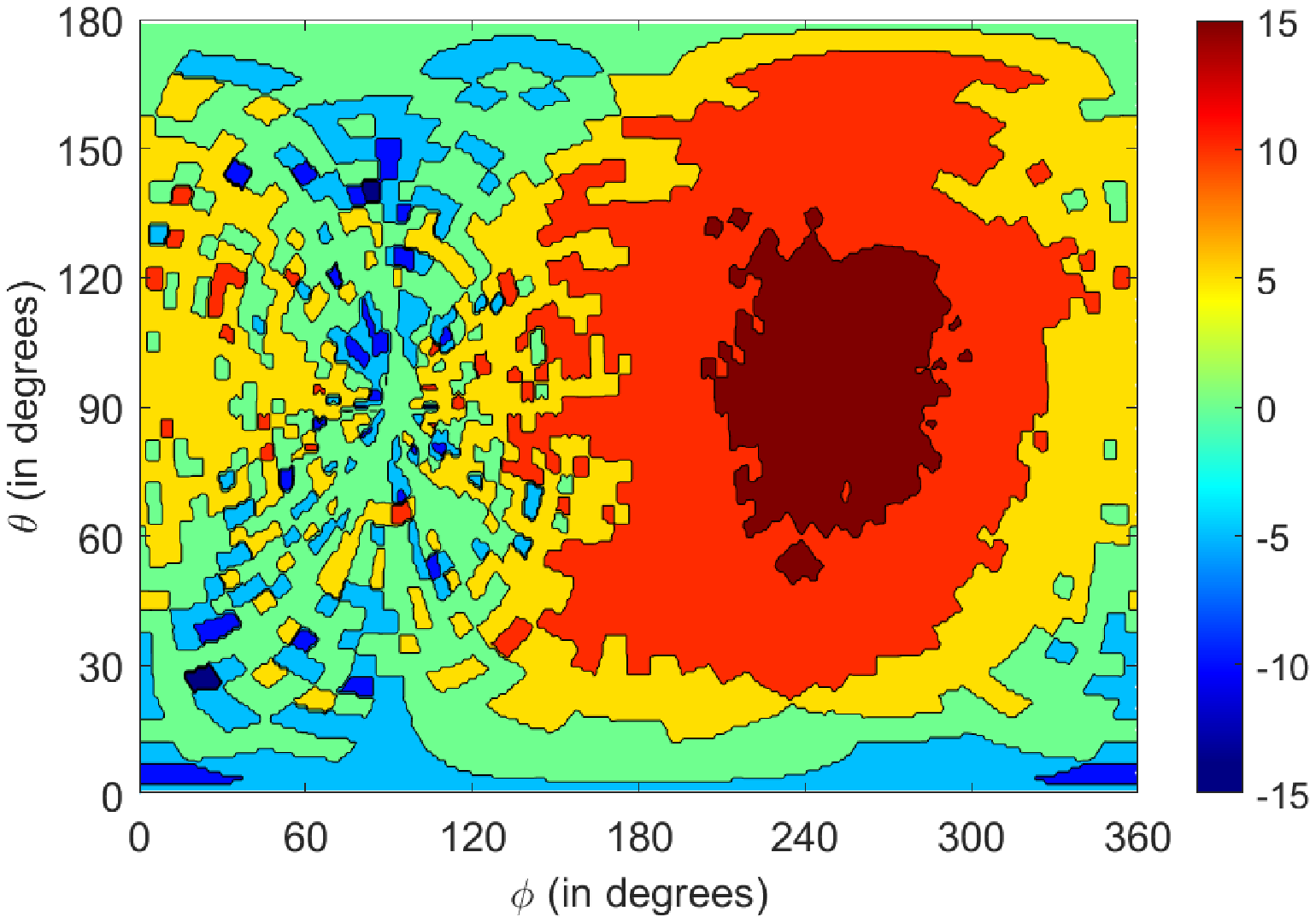}
&
\includegraphics[height=1.7in,width=1.5in]{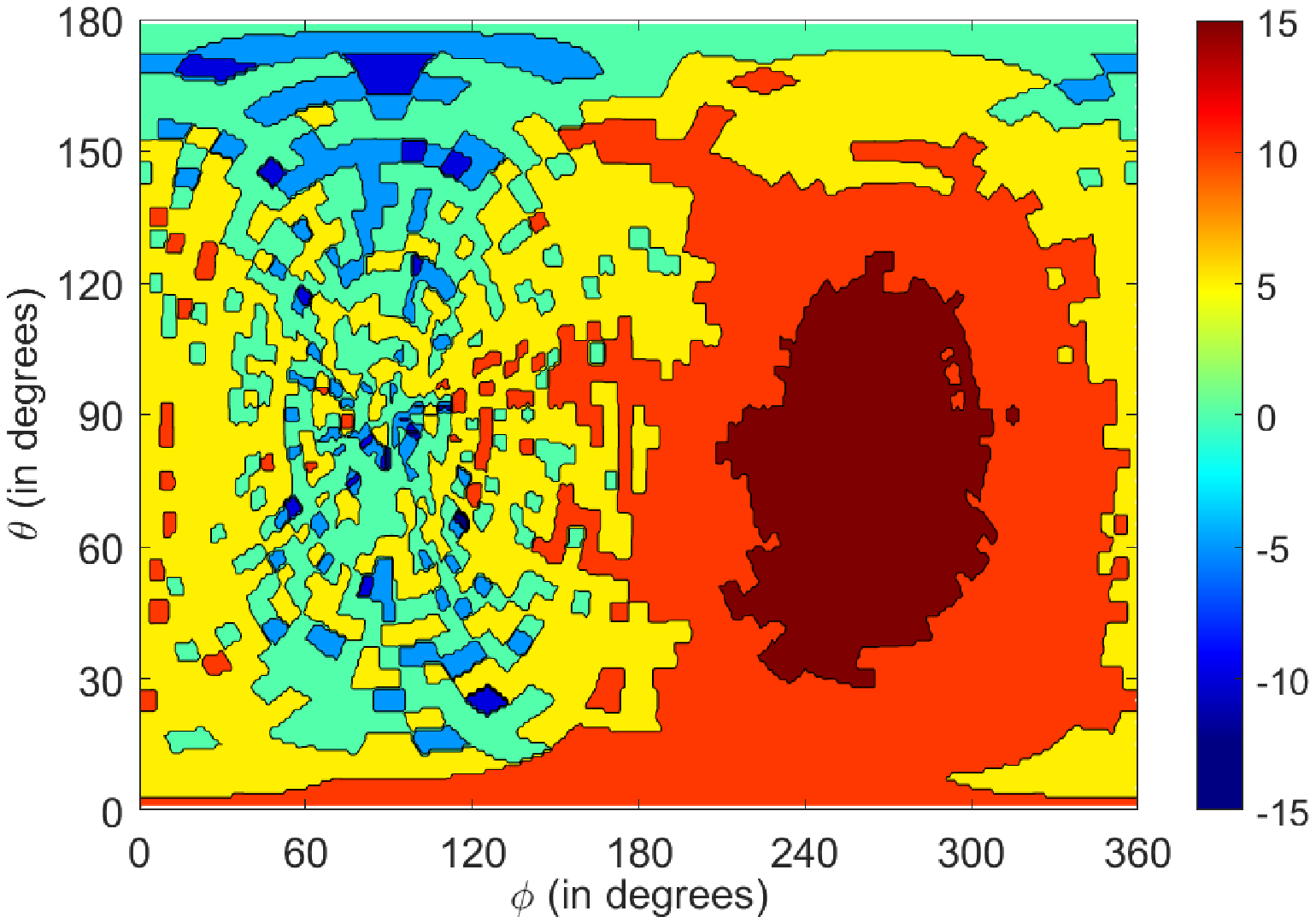}
\\
(a) & (b) & (c) & (d)
\\
\includegraphics[height=1.7in,width=1.5in]{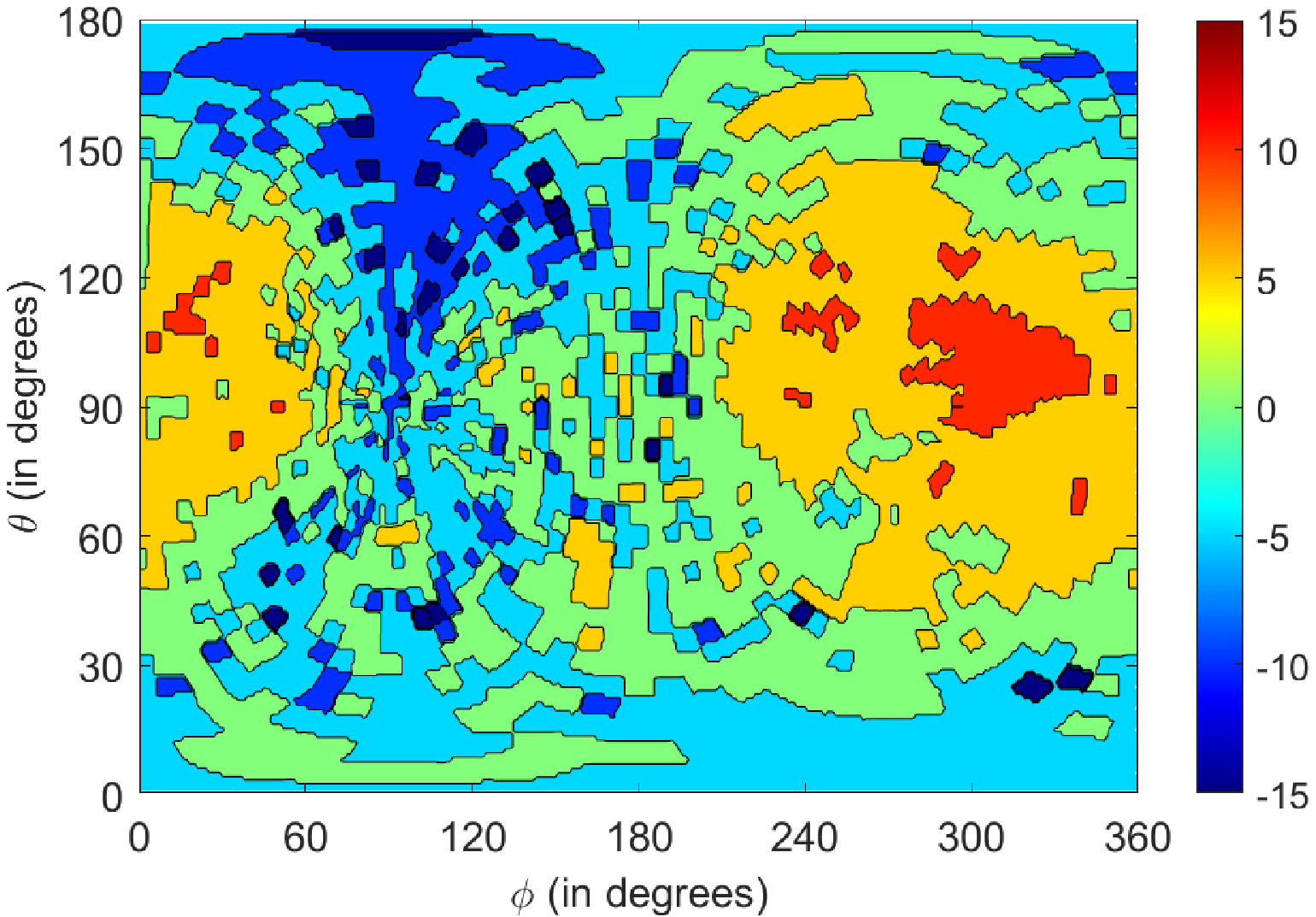}
&
\includegraphics[height=1.7in,width=1.6in]{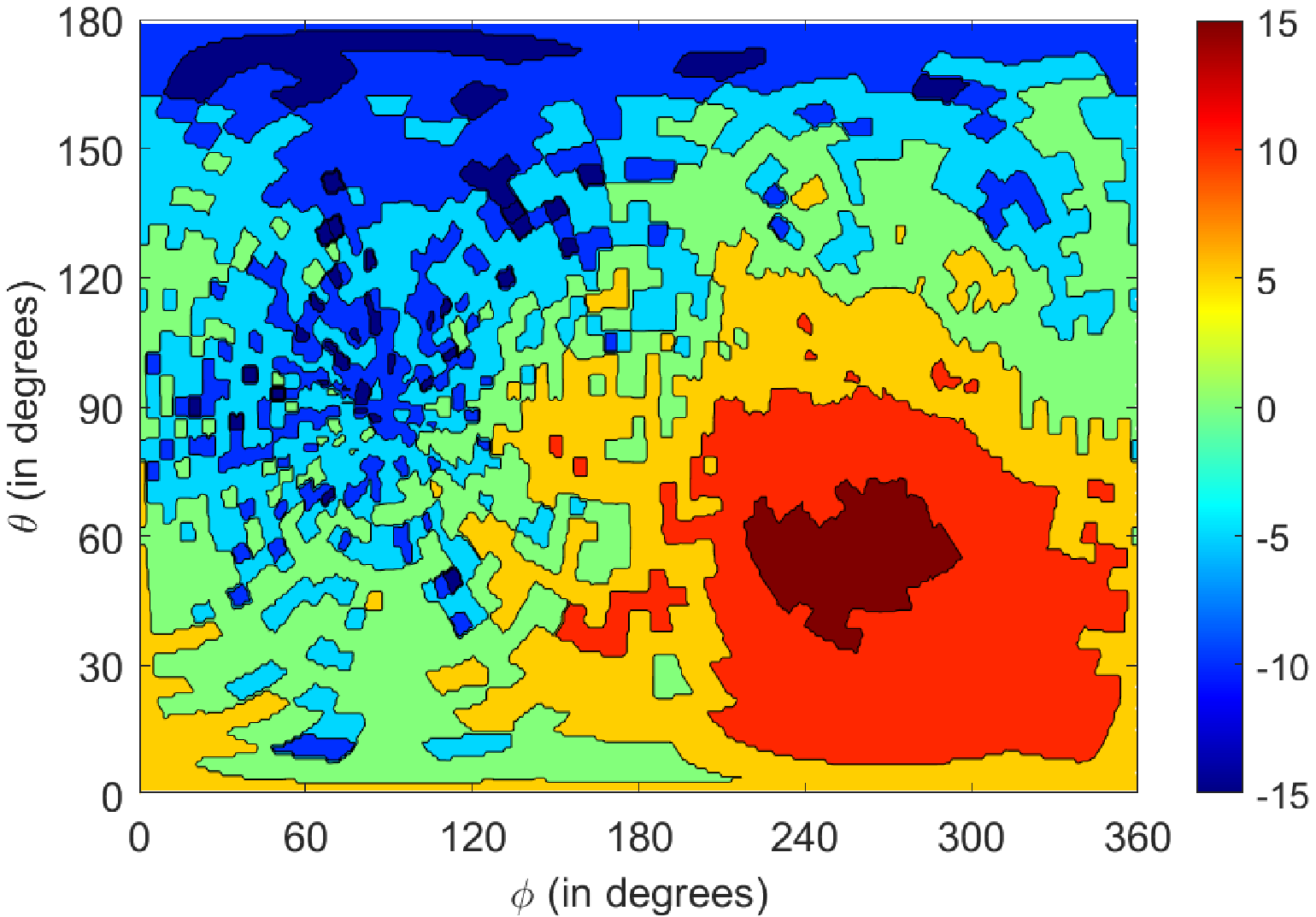}
&
\includegraphics[height=1.7in,width=1.6in]{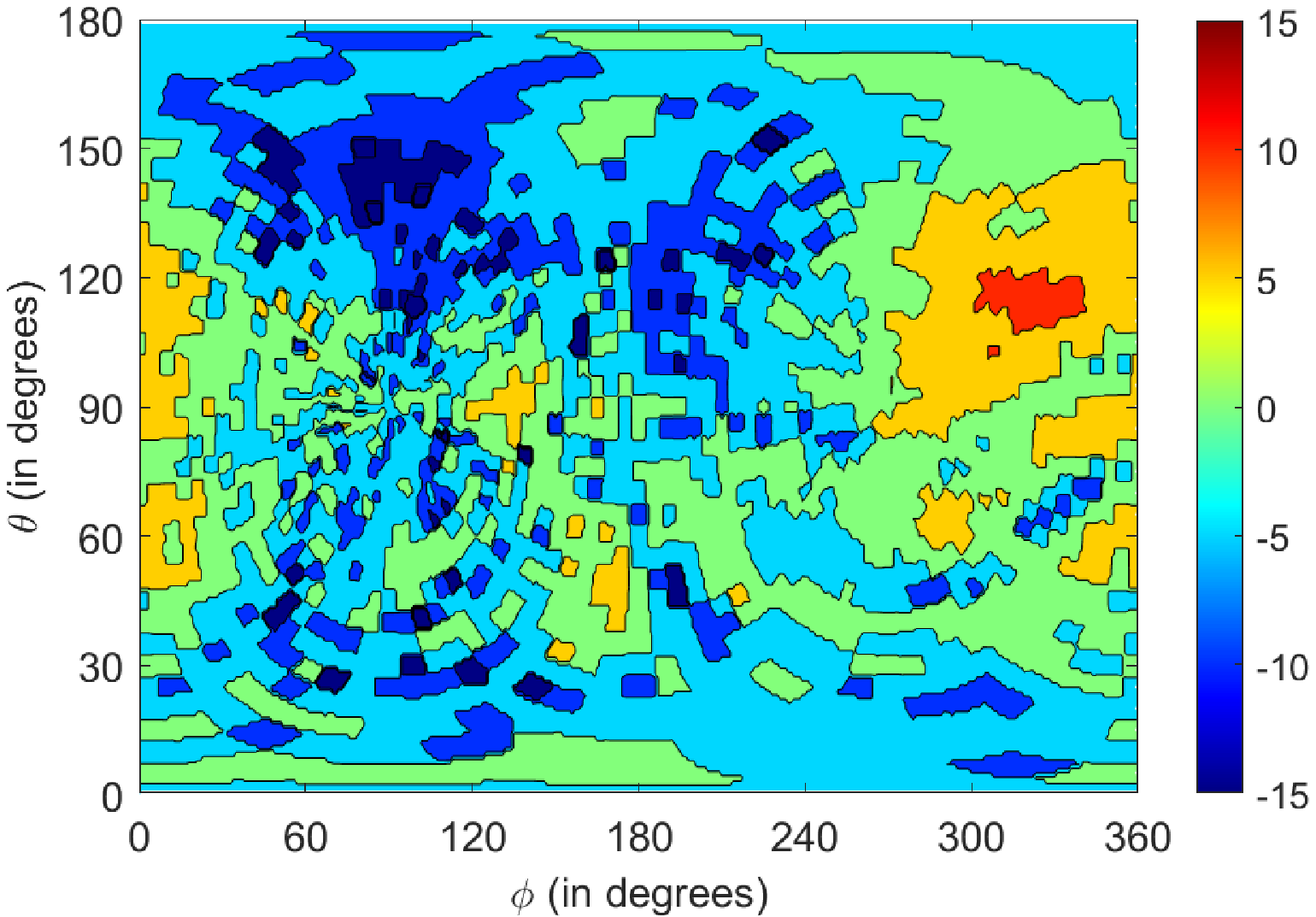}
&
\includegraphics[height=1.7in,width=1.5in]{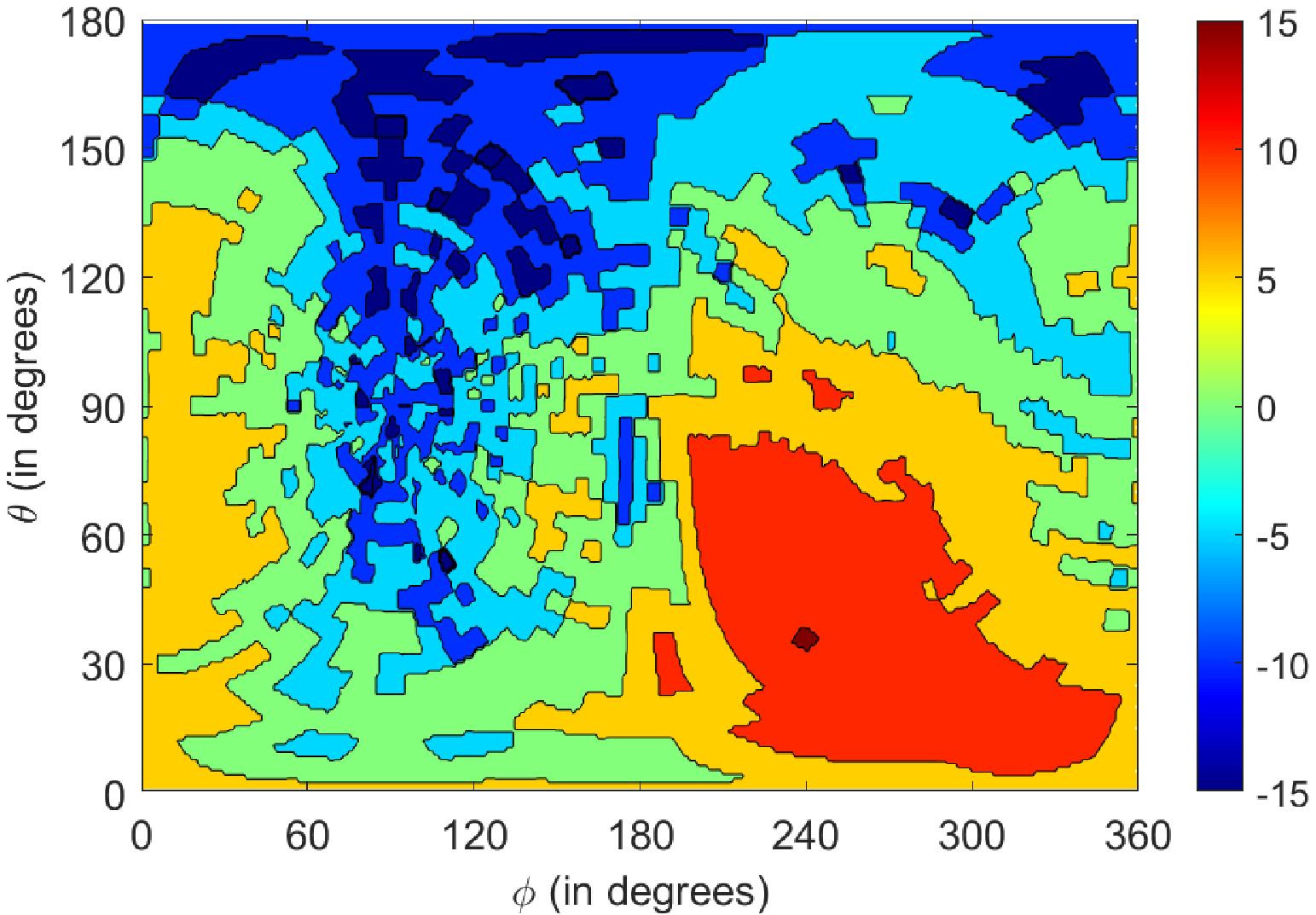}
\\
(e) & (f) & (g) & (h)
\\
\includegraphics[height=1.7in,width=1.5in]{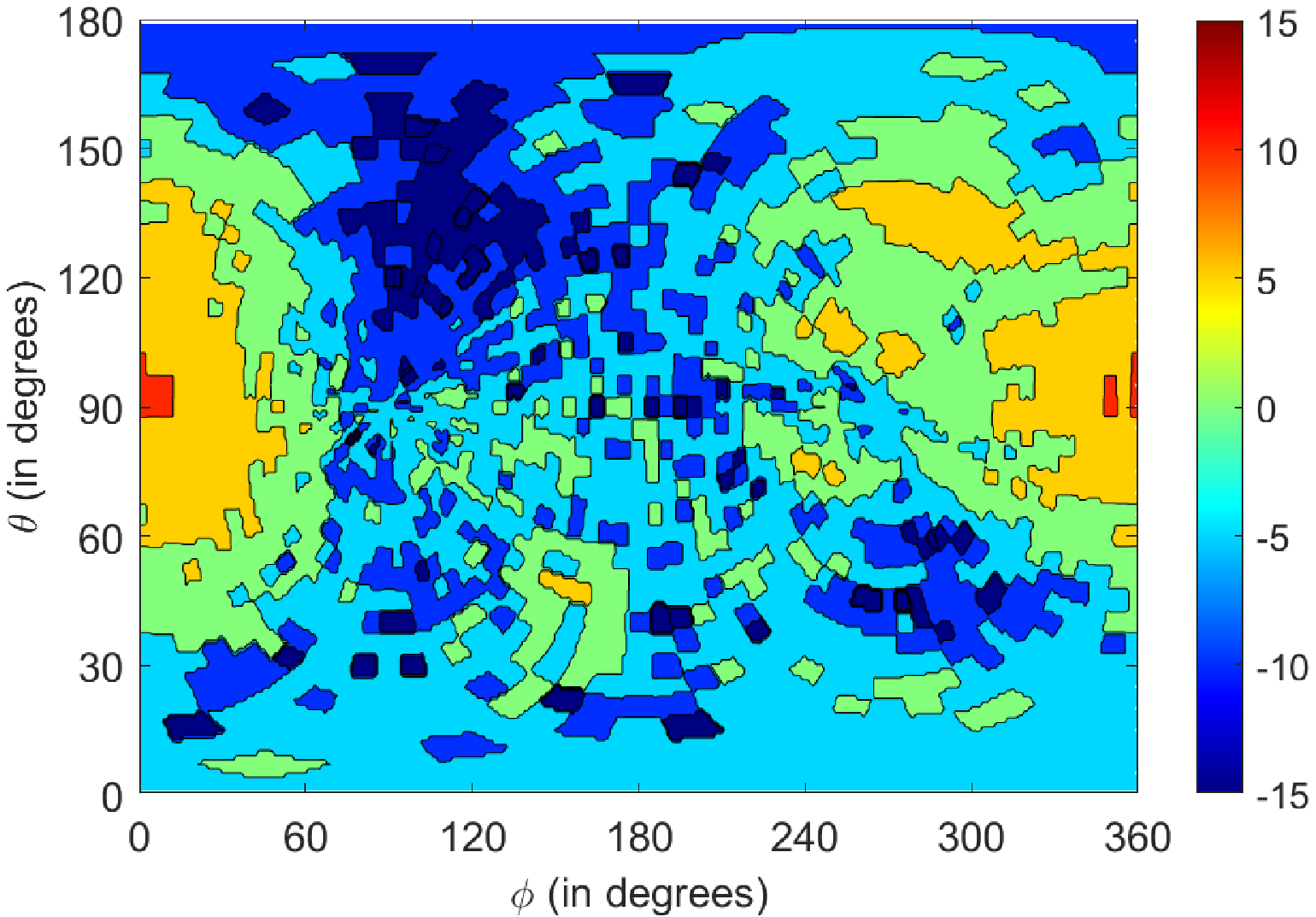}
&
\includegraphics[height=1.7in,width=1.6in]{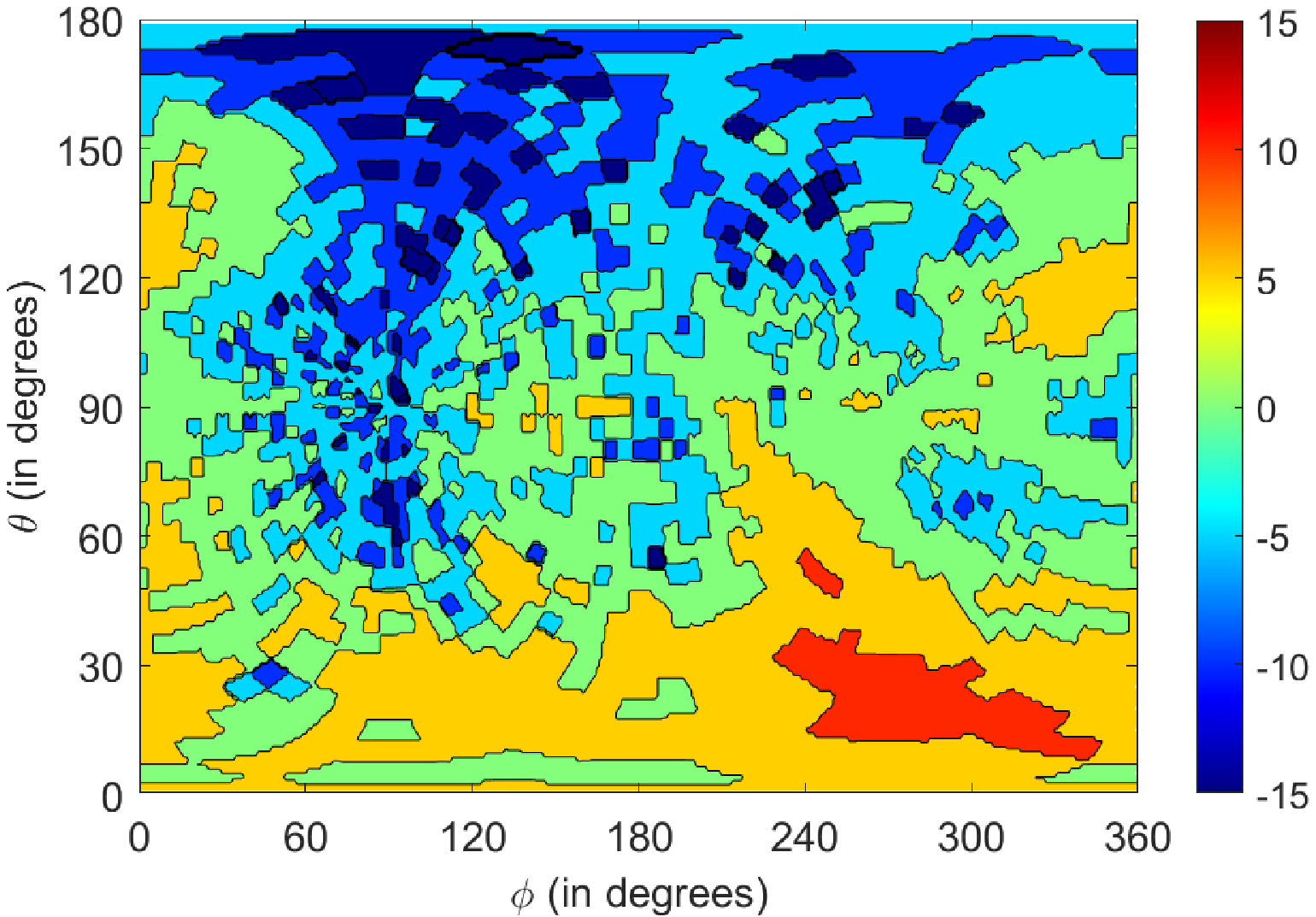}
&
\includegraphics[height=1.7in,width=1.6in]{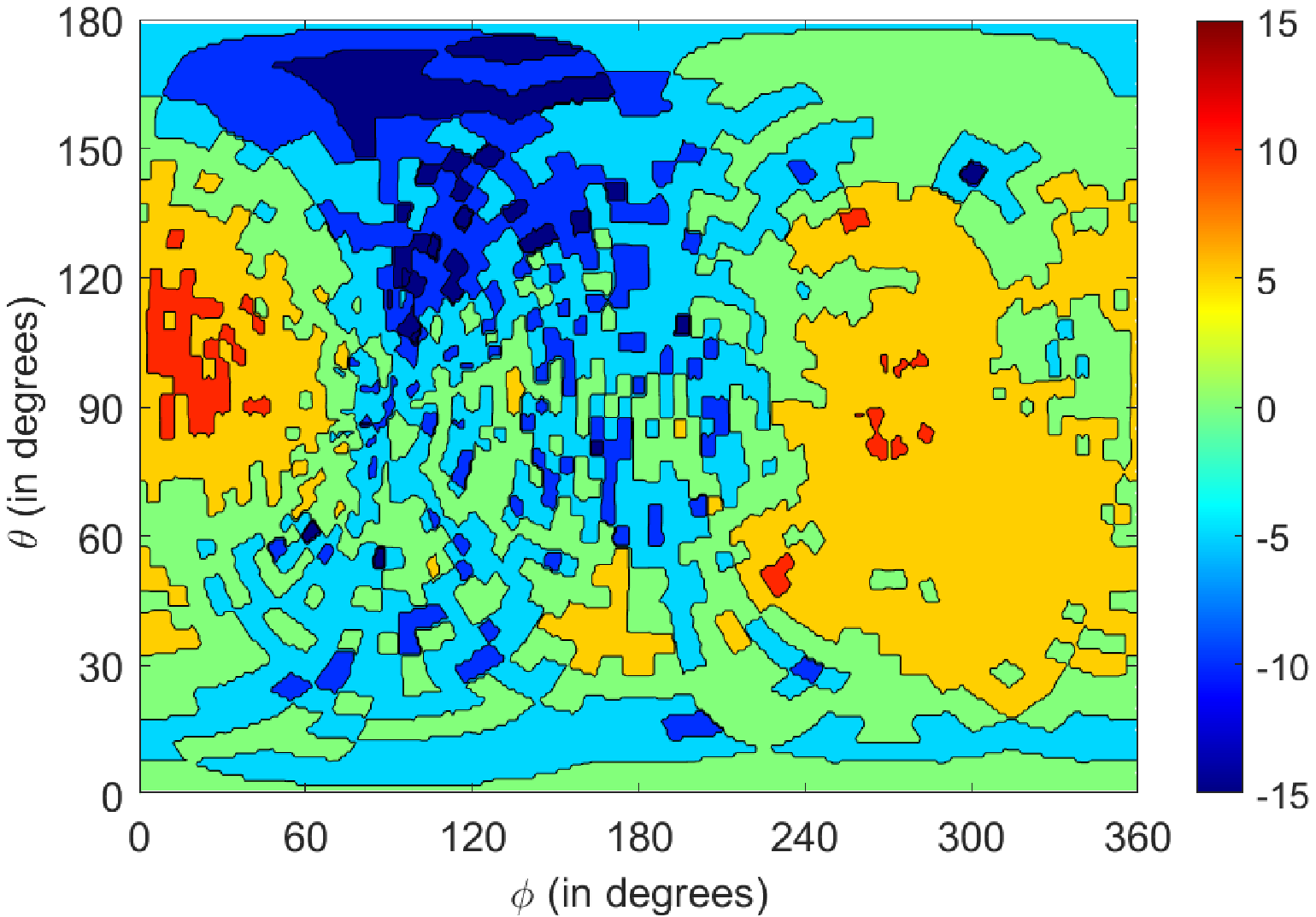}
&
\includegraphics[height=1.7in,width=1.5in]{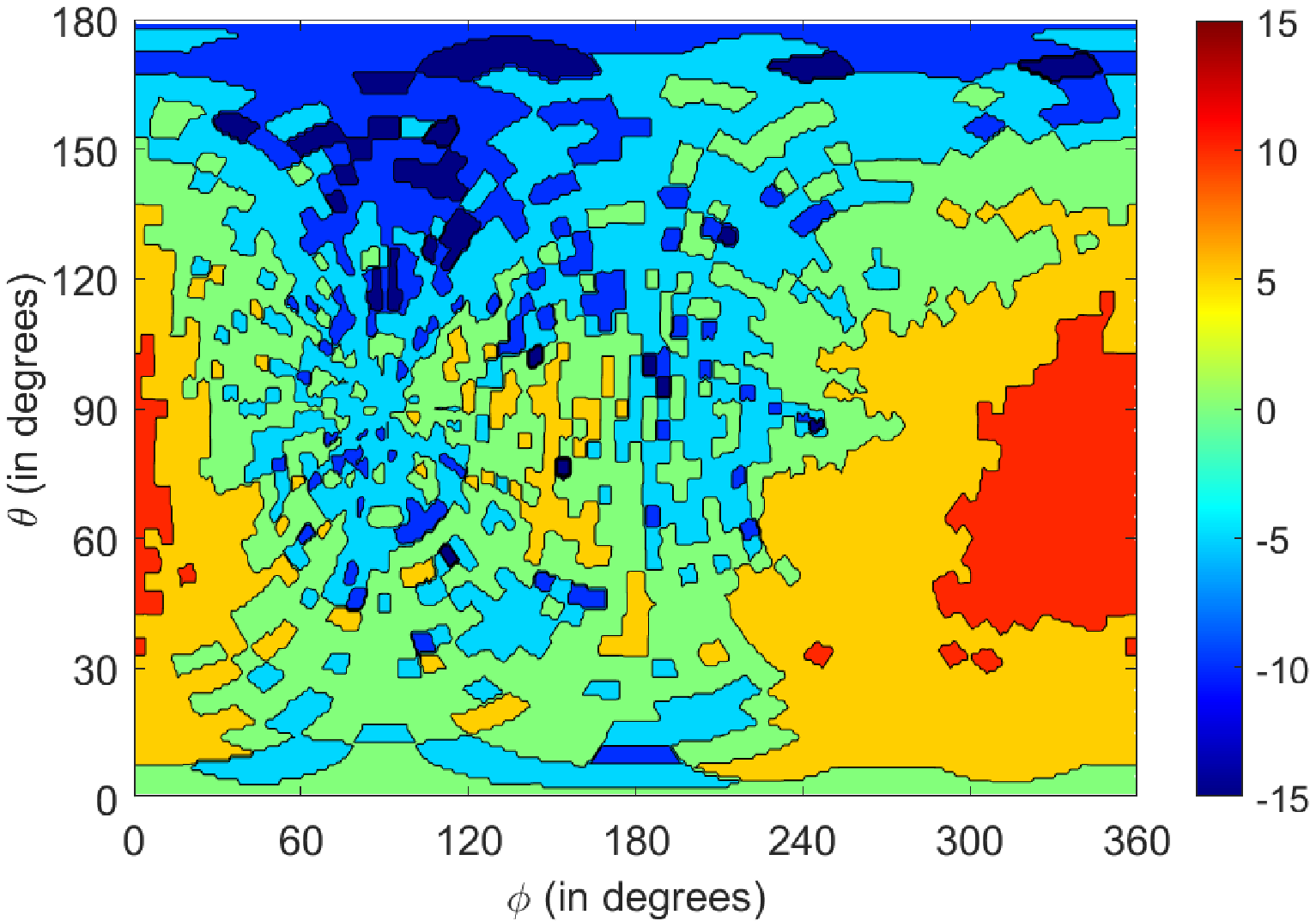}
\\
(i) & (j) & (k) & (l)
\end{tabular}
\caption{\label{fig_FS_Individual_ant_elements}
Contour plots over the sphere of elemental pattern of individual Antennas 0 to 3 
(a)-(d) in Freespace, (e)-(h) with $0$ mm air gap and one finger on the module, and
(i)-(l) with $0$ mm air gap and two fingers on the module.}
\end{center}
\end{figure*}

Electric field information (amplitudes and phases) that capture the elemental patterns of the four
individual antenna elements of the $4 \times 1$ array in Freespace are measured in the chamber.
These amplitudes and phases are recorded in $5^{\sf o}$ angular steps around the sphere (in azimuth
and elevation) in a global coordinate system, which translates to a finite but non-uniform precision 
in the coordinate system local to the UE. The amplitudes and phases are then extrapolated over a 
uniform grid of sample points and with this data, the contour plots of the elemental patterns of these 
four antenna elements in Freespace are plotted in Figs.~\ref{fig_FS_Individual_ant_elements}(a)-(d). 
These plots show that the individual antenna elemental gains peak around $\theta = 90^{\sf o}$ and $\phi = 270^{\sf o}$
(or -Y axis), as expected from the coordinate system presented in Fig.~\ref{fig_setup}. The elemental
patterns have good gains in $\approx 90^{\sf o} \times 120^{\sf o}$ of the sphere, 
which is also typical of antenna elements at millimeter wave carrier frequencies~\cite{vasanth_tcom2019}.

For the $0$ mm air gap scenario with one and two fingers on the antenna module,
Figs.~\ref{fig_FS_Individual_ant_elements}(e)-(h) and Figs.~\ref{fig_FS_Individual_ant_elements}(i)-(l)
illustrate the elemental patterns, 
respectively. Clearly, from these figures, we see significant signal strength distortion
(re-orientation of the peak direction as well as attenuation across the coverage region) for all
the antenna elements 
with loss in most directions, but occasional gains\footnote{Note that similar observations
of gains in some directions have also been made in~\cite{vasanth_blockage_2020} and~\cite{hejselbaek}.}
in some directions. From a visual point-of-view, signal distortions correspond to changes in regions
plotted as oceans of red in Figs.~\ref{fig_FS_Individual_ant_elements}(a)-(d) in Freespace to regions
plotted as orange, green and blue in Figs.~\ref{fig_FS_Individual_ant_elements}(e)-(h) and~(i)-(l), respectively.

\begin{figure*}[htb!]
\begin{center}
\begin{tabular}{cc}
\includegraphics[height=1.9in,width=2.7in]{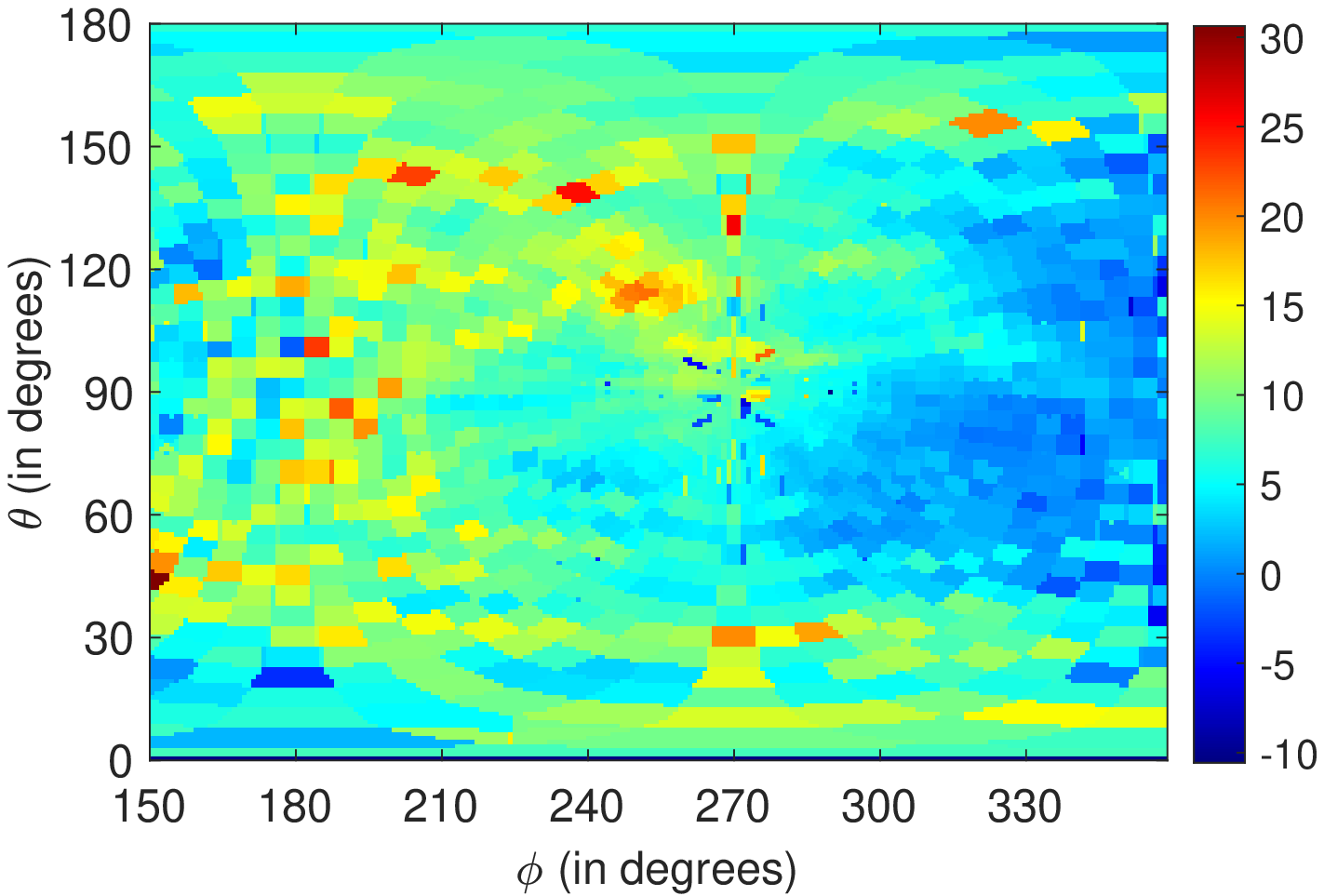}
&
\includegraphics[height=1.9in,width=2.7in]{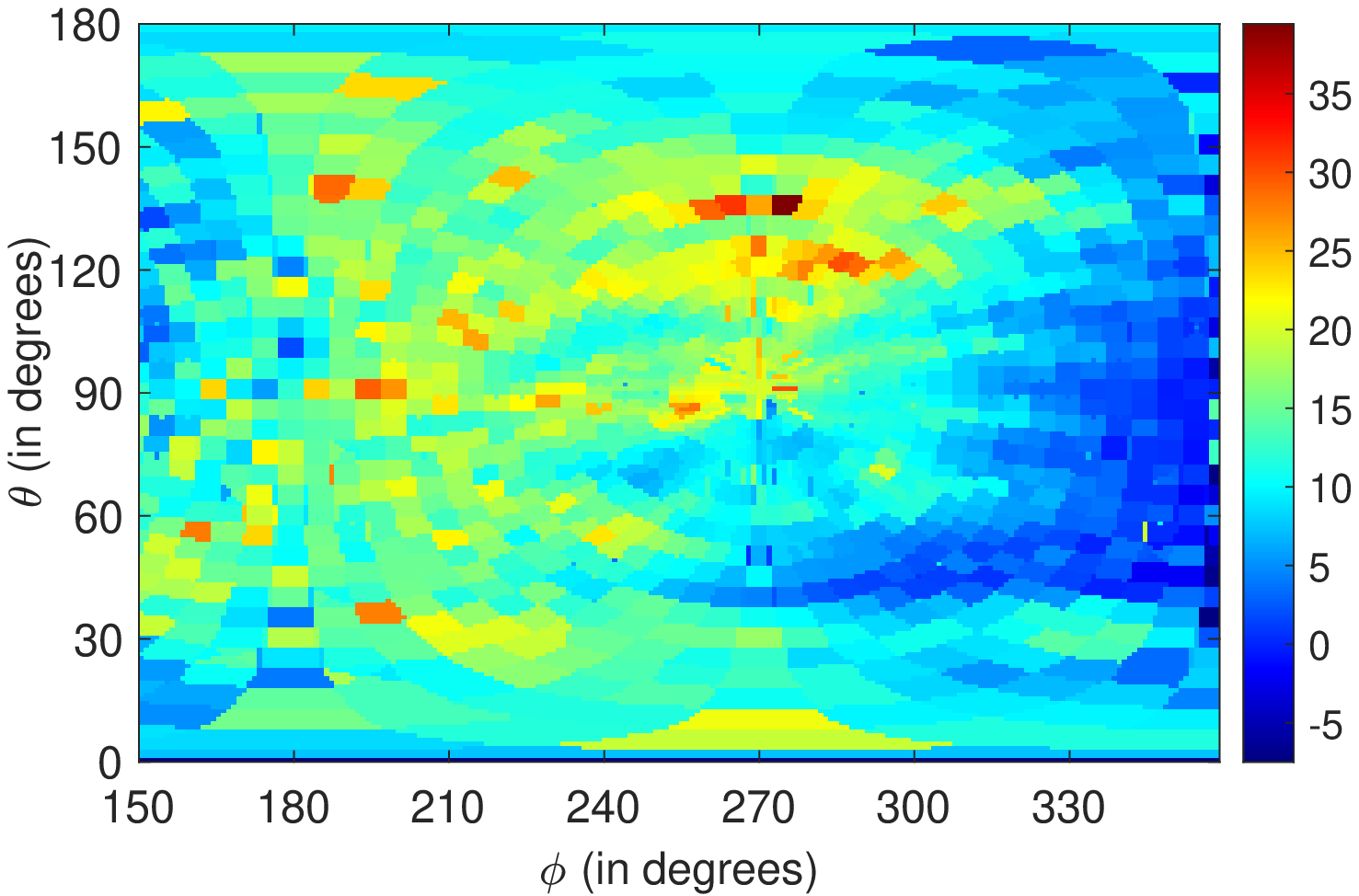}
\\
(a) & (b)
\end{tabular}
\caption{\label{fig_amp}
Amplitude distortion between the Freespace and blockage scenario with $0$ mm air gap corresponding 
to (a) one and (b) two fingers. }
\end{center}
\end{figure*}

Towards a quantitative study, let $E_{ {\sf free},i}(\theta,\phi)$ and $E_{ {\sf blockage},i}(\theta,\phi)$ denote the (complete) 
electric fields of the $i$-th antenna in the $(\theta, \phi)$ direction of the sphere in Freespace and with blockage, 
respectively. In a more careful comparative analysis of the impact of blockage on the amplitudes seen 
by the antenna elements, in Figs.~\ref{fig_amp}(a)-(b), we plot $10 \cdot \log_{10} \left( \left|
\frac{ E_{ {\sf free},  0}(\theta, \phi)  } { E_{ {\sf blockage},  0}(\theta, \phi) }
\right|^2 \right)$
for Antenna 0 over a {\em Region of Interest}\footnote{An RoI captures the region over the sphere 
where Freespace and/or blockage performance is relevant in terms of signal strengths observed. If 
the signal strength over a region is too poor for a viable link in both Freespace and blockage 
scenarios, that region is not part of the RoI. See~\cite{vasanth_blockage_2020} for a more detailed 
discussion on what an RoI entails and different mathematical descriptions/potential definitions.} 
(RoI) of the sphere in the $0$ mm air gap case with one and two fingers, respectively. The RoI chosen 
in this study is $[150^{\sf o}, \hsppp 360^{\sf o}) \times [0^{\sf o}, \hsppp 180^{\sf o})$ in 
azimuth and elevation, which covers $\approx 58.3\%$ of the sphere. From this plot, we observe that the presence 
of the hand either leads to significant losses or significant gains, 
or comparable amplitude responses relative to the Freespace scenario (blue regions in the plots).

In the one and two finger cases for the $0$ mm air gap, the finger(s) is/are approximately near Antenna 2
and Antennas 0 and 1, respectively. In the $1$ mm air gap, the finger(s) is/are approximately near
Antenna 3 and Antennas 0 and to the left of Antenna 0, respectively. This is captured by the significant 
distortion in the elemental patterns in these settings. To accurately quantify the blockage losses with 
different antenna elements in contrast to the na{\"i}ve RoI as in Fig.~\ref{fig_amp}, we define a RoI over 
the sphere associated with the $i$-th antenna as follows:
\begin{eqnarray}
{\sf RoI}_i = \Big\{ (\theta, \phi) \hsppp : \hsppp G_{ {\sf free},i}(\theta,\phi) \geq
G_1 \hspp {\sf or} \hspp G_{ {\sf blockage}, i}(\theta, \phi) \geq G_2
\Big\}
\nonumber
\end{eqnarray}
where
\begin{eqnarray}
G_{ {\sf free},i}(\theta,\phi) & = & 10 \cdot \log_{10} \left( |E_{ {\sf free},i}(\theta,\phi)|^2
\right) {\hspace{0.05in}} {\sf and} \nonumber \\
G_{ {\sf blockage}, i}(\theta, \phi) & = &
10 \cdot \log_{10} \left( |E_{ {\sf blockage},i}(\theta,\phi) |^2  \right)
\nonumber
\end{eqnarray}
denote the gains (in dB) of the $i$-th antenna in Freespace and with blockage, respectively. The 
thresholds $G_1$ are $G_2$ are defined appropriately (see~\cite{vasanth_blockage_2020} for a 
discussion on good choices of thresholds). In general, a small/narrow definition of the RoI does 
not capture the impact of hand reflections, whereas a large/broad definition incorporates poor link 
budget regions in analysis. To optimize these tradeoffs, in this paper, we use the values 
$G_1 = 7.5$ dB and $G_2 = 2.5$ dB since $55$-$70\%$ of the sphere is included in ${\sf RoI}_i$ 
with these choices for both the $0$ mm and $1$ mm air gap cases, which is a good compromise (see 
Table~\ref{table_area}).

\begin{table}[htb!]
\caption{Coverage Area Properties with $G_1 = 7.5 {\hspace{0.05in}} {\mathrm{dB}}$ 
and $G_2 = 2.5 {\hspace{0.05in}} {\mathrm{dB}}$}
\label{table_area}
\begin{center}
\begin{tabular}{|c|| c|c|c|c|}
\hline 
Antenna index ($\rightarrow$) & 0 & 1 & 2 & 3 \\ 
\hline 
$\max_{\theta,\phi}  G_{ {\sf free},i}(\theta,\phi)$ (in dBi)
& 17.8 & 17.1 & 18.1 & 17.8 
\\ 
\hline 
\end{tabular}
\newline
\vspace*{0.05in}
\newline
\begin{tabular}{|c|| c|c|c|c|| c|c|c|c| }
\hline
Antenna index ($\rightarrow$) & 0 & 1 & 2 & 3 & 0 & 1 & 2 & 3 \\
\hline
& \multicolumn{4}{c||}{0 mm air gap, 1 finger} & \multicolumn{4}{c|}{0 mm air gap, 2 fingers} \\
\hline
$\max_{\theta,\phi}  G_{ {\sf blockage},i}(\theta,\phi)$ (in dBi)
& 12.7 & 16.6 & 11.4 & 15.1 & 11.2 & 11.6 & 12.0 & 14.0 \\ \hline
Area of $\left\{ G_{ {\sf free},i}(\theta,\phi) \geq G_1  {\hspace{0.03in}}
{\sf or} {\hspace{0.03in}} G_{ {\sf blockage},i}(\theta,\phi) \geq G_2 \right\}$ (in $\%$)
& 60.1 & 64.2 & 56.2 & 67.5  & 57.6 & 68.3 & 62.9 & 67.6  \\
\hline 
\hline
& \multicolumn{4}{c||}{1 mm air gap, 1 finger} & \multicolumn{4}{c|}{1 mm air gap, 2 fingers} \\
\hline
$\max_{\theta,\phi}  G_{ {\sf blockage},i}(\theta,\phi)$ (in dBi)
& 16.3 & 15.1 & 16.8 & 13.9 & 12.7 & 14.9 & 13.6 & 13.7 \\ \hline
Area of $\left\{ G_{ {\sf free},i}(\theta,\phi) \geq G_1  {\hspace{0.03in}}
{\sf or} {\hspace{0.03in}} G_{ {\sf blockage},i}(\theta,\phi) \geq G_2 \right\}$ (in $\%$)
& 66.6 & 72.0 & 66.1 & 74.3  & 59.5 & 69.8 & 65.3 & 68.4  \\
\hline
\hline
\end{tabular}
\end{center}
\end{table}

\begin{figure}[htb!]
\begin{center}
\begin{tabular}{cc}
\includegraphics[height=2.0in,width=2.7in]{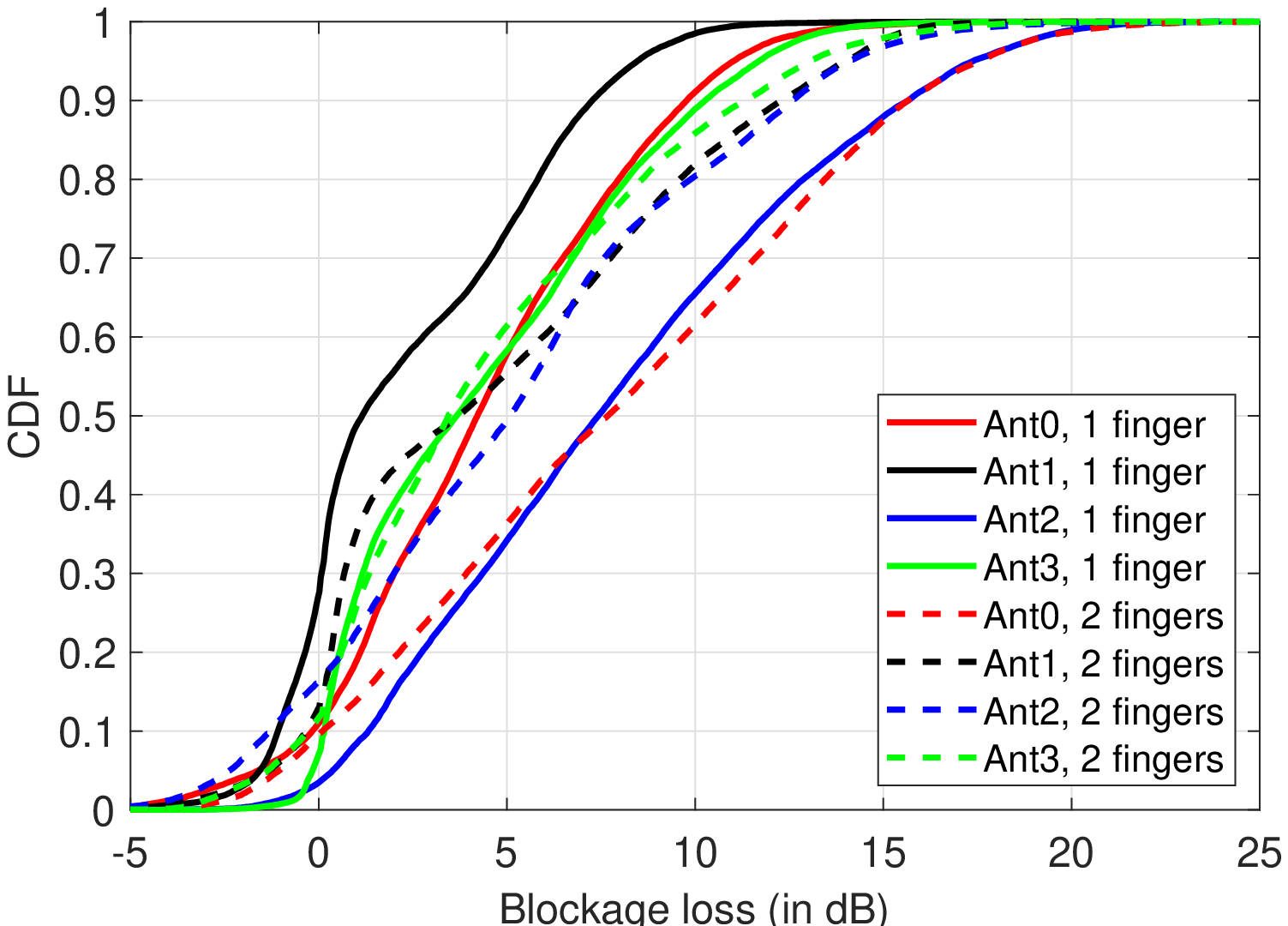}
&
\includegraphics[height=2.0in,width=2.7in]{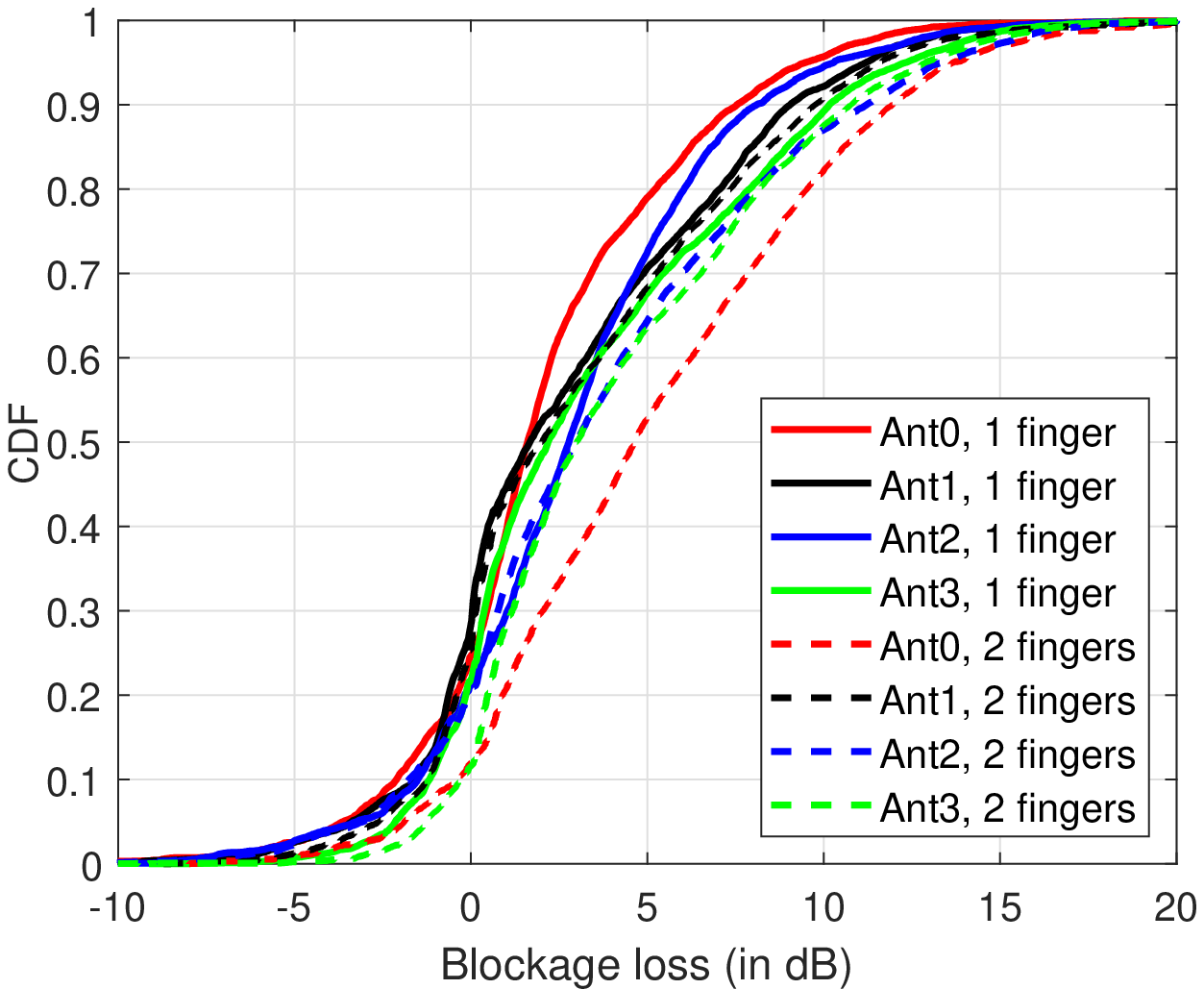}
\\
(a) & (b)
\end{tabular}
\caption{\label{fig_single_ant_elem_losses}
CDF of blockage losses with individual antenna elements for the (a) $0$ mm and (b) $1$ mm air gap scenarios
with $1$ and $2$ fingers on the antenna module.}
\end{center}
\end{figure}

\begin{table}[htb!]
\caption{Blockage Loss Statistics with Individual Antenna Elements}
\label{table_ind_elements}
\begin{center}
\begin{tabular}{
|c|| c|c|c|c|| c|c|c|c| }
\hline
& \multicolumn{4}{c||}{0 mm air gap, 1 finger} & \multicolumn{4}{c|}{0 mm air gap, 2 fingers} \\
\hline
Antenna index ($\rightarrow$) & 0 & 1 & 2 & 3 & 0 & 1 & 2 & 3 \\
\hline
Mean (in dB) & 4.3 & 2.5 & 7.5 & 4.5 & 7.6 & 4.7 & 5.1 & 4.6 \\
\hline
Std. dev (in dB) & 4.4 & 3.8 & 5.8 & 4.3 & 6.2 & 5.3 & 5.5 & 4.8 \\
\hline \hline
& \multicolumn{4}{c||}{1 mm air gap, 1 finger} & \multicolumn{4}{c|}{1 mm air gap, 2 fingers} \\
\hline
Antenna index ($\rightarrow$) & 0 & 1 & 2 & 3 & 0 & 1 & 2 & 3 \\
\hline
Mean (in dB) & 2.2 & 2.8 & 3.0 & 3.5 & 5.3 & 3.2 & 3.8 & 4.3 \\
\hline
Std. dev (in dB) & 3.9 & 4.6 & 4.1 & 4.6 & 4.9 & 4.6 & 5.1 & 4.5 \\
\hline \hline
\end{tabular}
\end{center}
\end{table}

With these choices, Figs.~\ref{fig_single_ant_elem_losses}(a) and~(b) capture the cumulative distribution
functions (CDFs) of losses with the individual antenna elements (that is, $G_{ {\sf free}, i}(\theta, \phi)
- G_{ {\sf blockage}, i}(\theta, \phi)$ with $(\theta, \phi) \in {\sf RoI}_i$) for the $0$ mm and $1$ mm
air gap cases considered here. In general, we note that the losses seen with two fingers are more than those 
seen in the one finger case. Also, the losses seen with the $1$ mm air gap are smaller than when the hand 
touches the antenna module ($0$ mm air gap). Both these observations are intuitive and obvious as two fingers 
obstruct more coverage than one finger, and the $1$ mm air gap allows creeping of electromagnetic waves and thus 
better energy reception than with fingers on the antenna module.
These observations are also captured in Table~\ref{table_ind_elements} which reports the mean and
standard deviation of blockage losses over ${\sf RoI}_i$ for the different antenna elements in these settings.

\section{Impact of Beamforming on Blockage}
\label{sec4}

\subsection{Optimal/Infinite-Precision Beamforming}
\label{sec4a}

While the studies in Sec.~\ref{sec3} considered the amplitude distortions seen at an individual antenna
element level (which are of broader interest in initial link acquisition~\cite{raghavan_jstsp,vasanth_tcom2019}), for 
peak performance,
the $4 \times 1$ array is used with analog beamforming. To understand the implications of blockage
in a practical context, we first consider the optimal beamforming solution along the direction
$(\theta, \phi)$ 
where $\bullet$ is used to denote $\bullet \in \{ {\sf free}, {\sf blockage} \}$ and
$^{\star}$ denotes the complex conjugate operation. This solution is given as 
\begin{eqnarray}
{\bf G}_{ {\sf opt}, \hsppp \bullet} (\theta, \phi) =
\max_ {w_i \hsppp : \hsppp \sum_{i} |w_i|^2 \leq 1}
10 \cdot \log_{10} \left( \left| \sum_{i=1}^N w_i^{\star} E_{ \bullet, \hsppp i}(\theta, \phi) \right|^2
\right)
\nonumber
\end{eqnarray}
which can be seen to be the maximum ratio combining (MRC) solution with 
\begin{eqnarray}
w_i \Big|_{\sf opt} = \frac{ E_{ \bullet, \hsppp i}(\theta, \phi) }
{ \sqrt{ \sum_{i = 1}^N |E_{ \bullet, \hsppp i}(\theta, \phi)|^2 } }
\nonumber
\end{eqnarray}
resulting in
\begin{eqnarray}
{\bf G}_{ {\sf opt}, \hsppp \bullet} (\theta, \phi) = 10 \cdot \log_{10} \left(
\sum_{i = 1}^N |E_{ \bullet, \hsppp i}(\theta, \phi)|^2 \right).
\nonumber
\end{eqnarray}
It is important to note that the above solution requires infinite-precision phase and amplitude
control (that is, an infinite number of beams in the codebook) and is hence not practical in 
implementations. In this context, the main purpose of this solution is only to benchmark the 
performance of more practical codebook-based schemes relative to an upper bound on performance.

\begin{figure*}[htb!]
\begin{center}
\begin{tabular}{ccc}
\includegraphics[height=1.9in,width=2.1in]{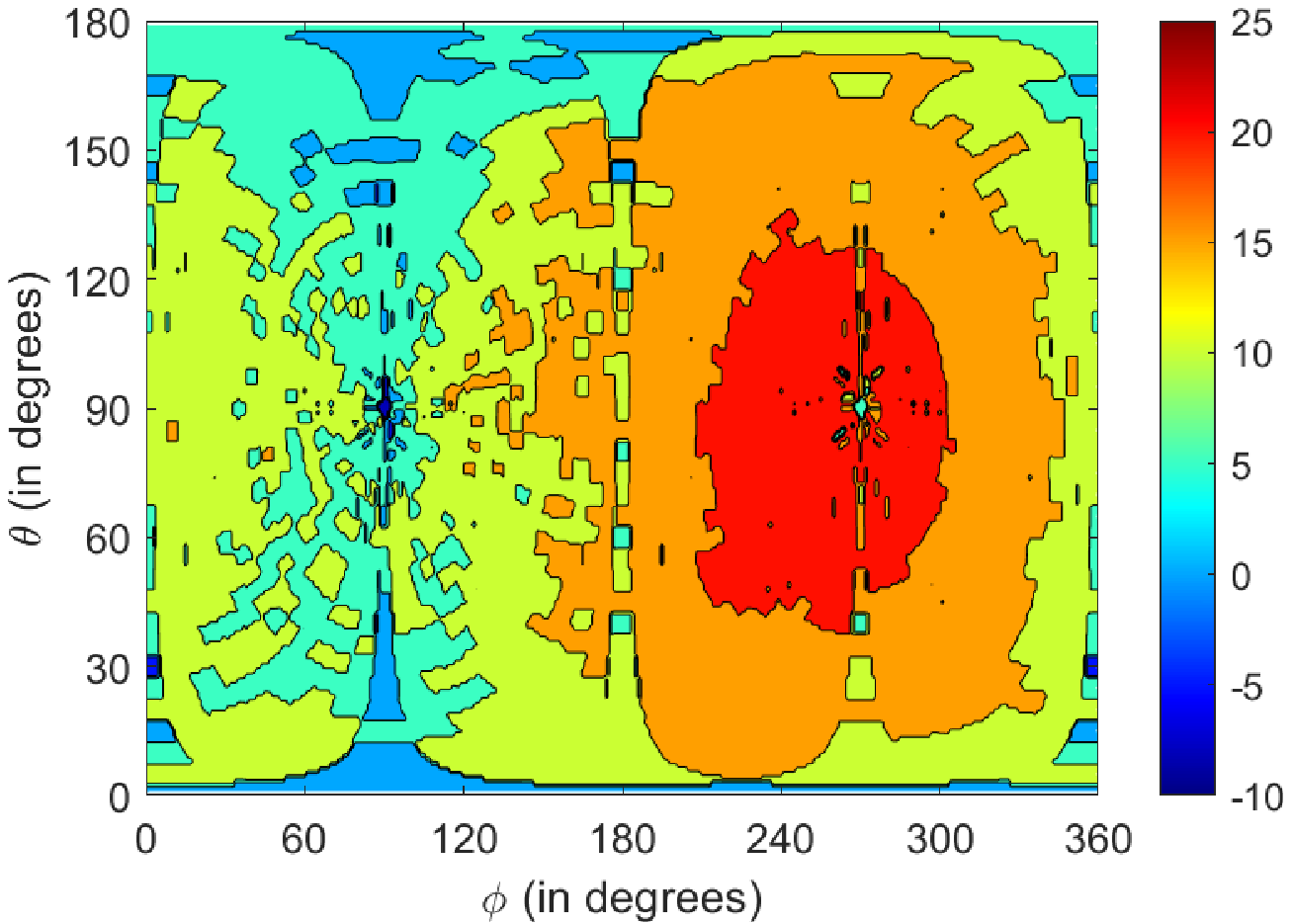}
&
\includegraphics[height=1.9in,width=2.1in]{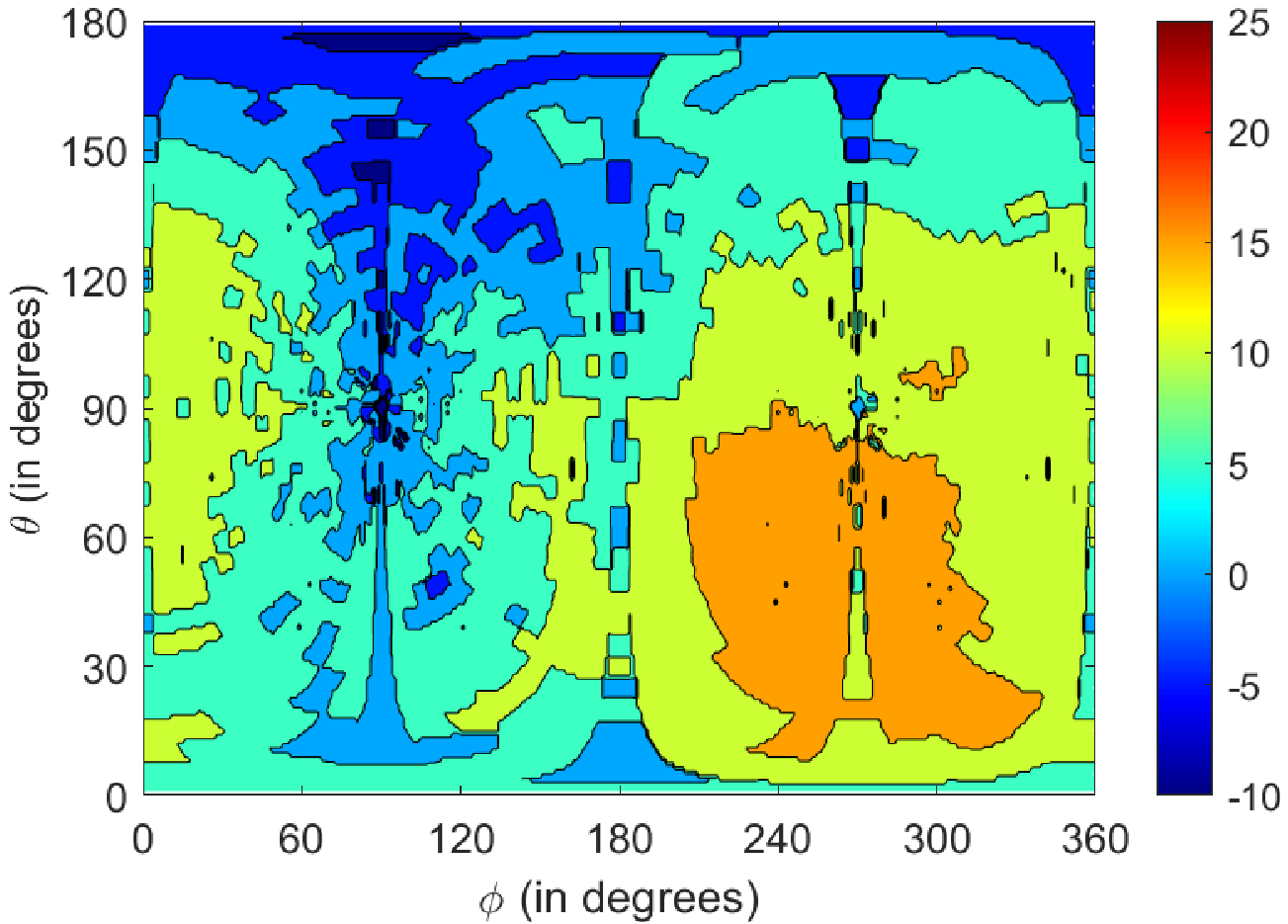}
&
\includegraphics[height=1.9in,width=2.1in]{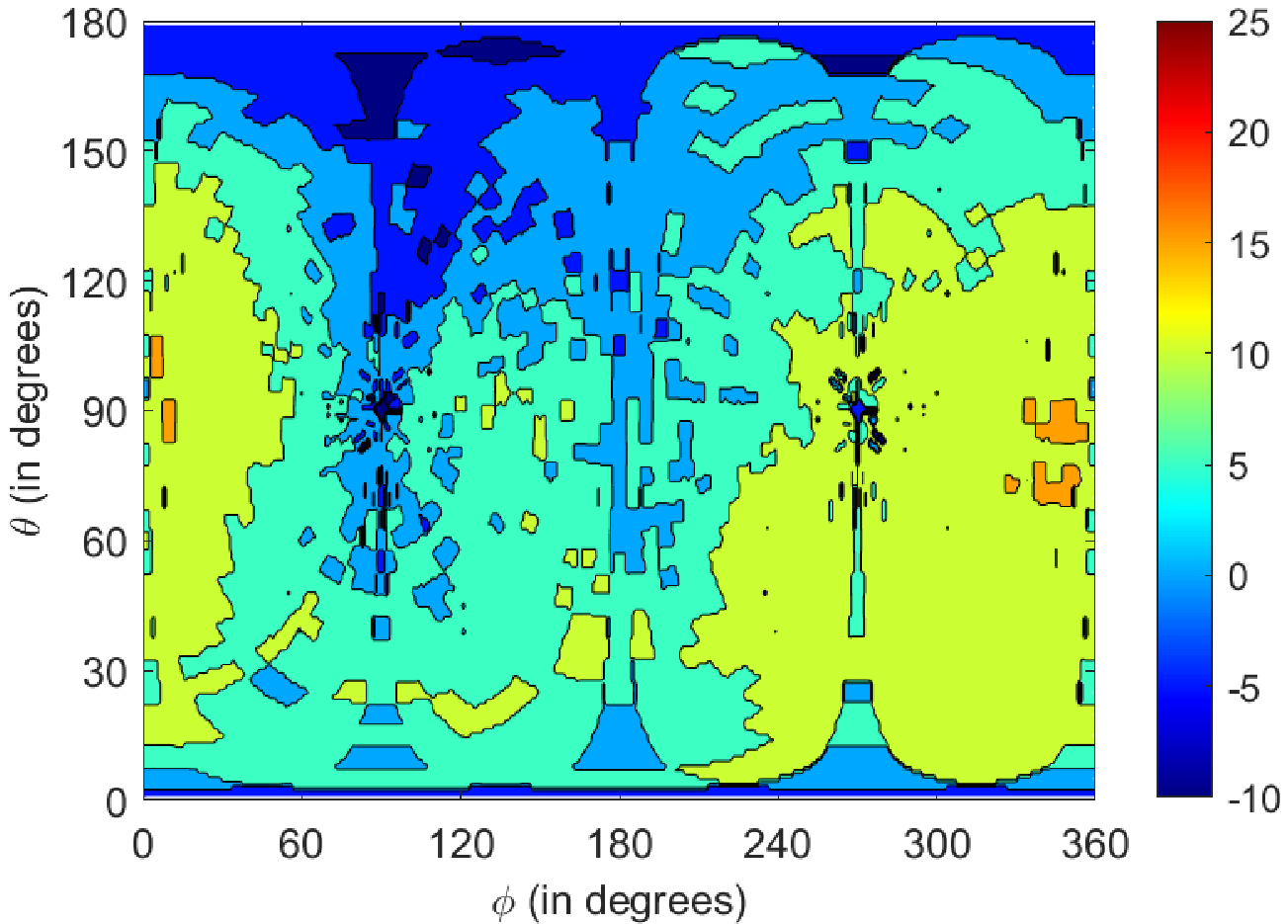}
\\
(a) & (b) & (c)
\\
\includegraphics[height=1.9in,width=2.1in]{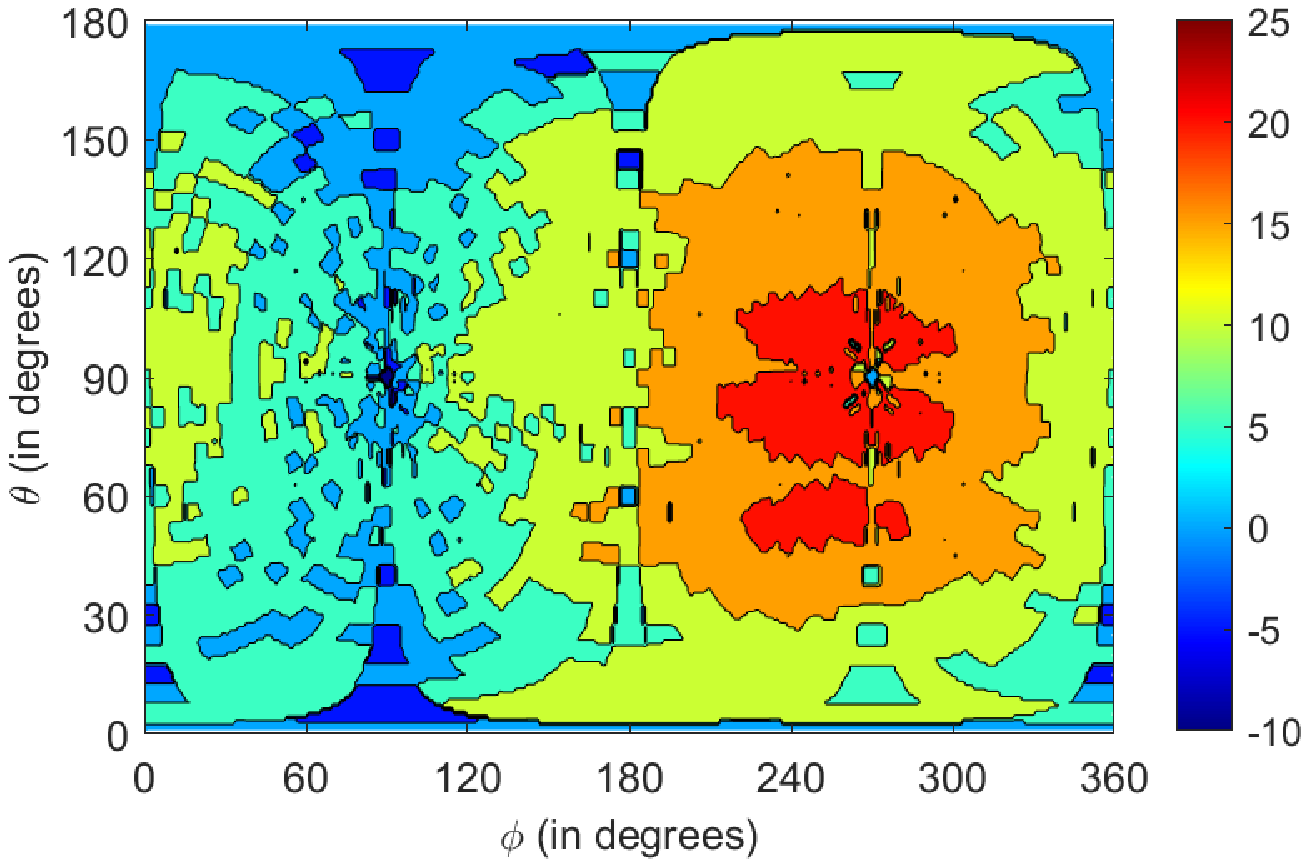}
&
\includegraphics[height=1.9in,width=2.1in]{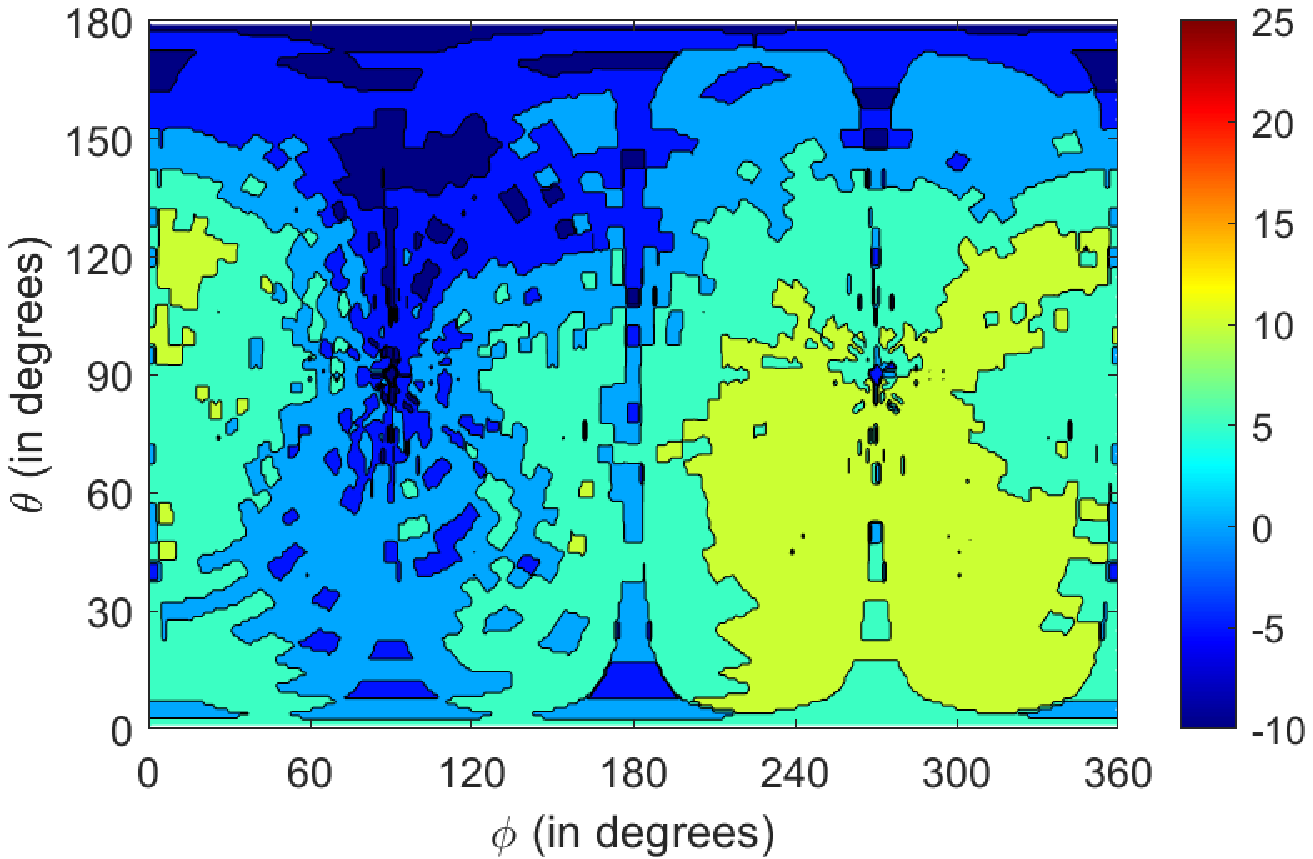}
&
\includegraphics[height=1.9in,width=2.1in]{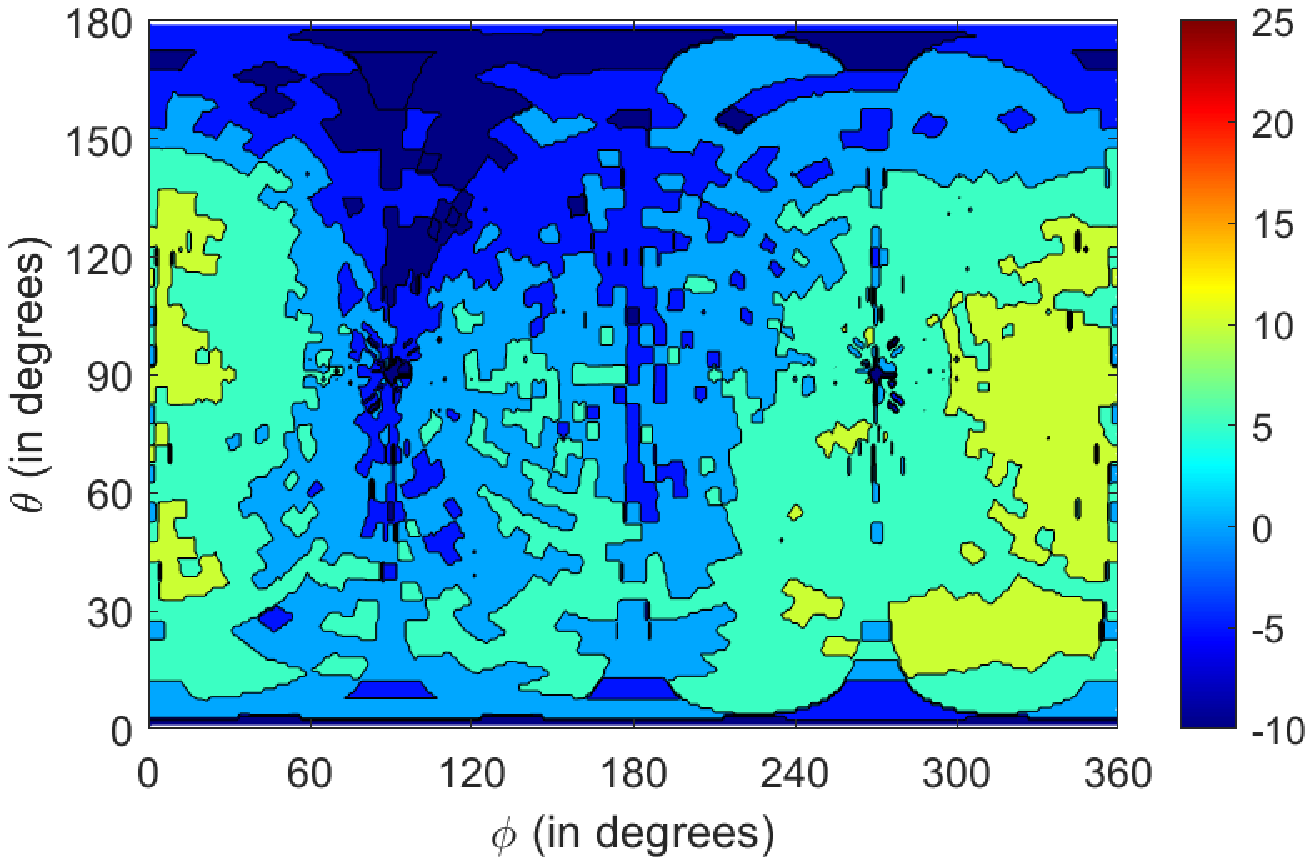}
\\
(d) & (e) & (f)
\end{tabular}
\caption{\label{fig_optbf}
Gain with (a)-(c) optimal and (d)-(f) codebook-based
beamforming over the sphere in Freespace, and $0$ mm air gap with one and two fingers, respectively.
}
\end{center}
\end{figure*}

Contour plots capturing the optimal beamforming gain over the sphere in Freespace and with one or two
fingers in the $0$ mm air gap case are plotted in Figs.~\ref{fig_optbf}(a)-(c). These plots show that
hand blockage can lead to significant performance degradation over a good part of the coverage area of
the antenna module. To quantify these gains, the CDFs of the optimal beamforming gain are plotted for
the different scenarios in Fig.~\ref{fig_opt}(a) over the RoI of $(\theta, \phi) = \left(
150^{\sf o}, \hsppp 360^{\sf o} \right) \times \left( 0^{\sf o} , \hsppp 180^{\sf o} \right)$. The median beamforming
gain in Freespace is $11.5$ dB, whereas in the $0$ mm air gap scenario, the median gains are 
$7.4$ dB and $5.6$ dB (one and two fingers). For the $1$ mm air gap scenario, the median gains are 
$8.0$ dB and $6.9$ dB, respectively. These numbers show that a $3.5$ to $6$ dB median loss is seen with 
blockage, consistent with trends on loose hand grips reported in~\cite{vasanth_blockage_2020}. To be 
precise about blockage losses, in Fig.~\ref{fig_opt}(b), we compare ${\bf G}_{ {\sf opt}, \hsppp 
{\sf free} } (\theta, \phi)$ with ${\bf G}_{ {\sf opt}, \hsppp {\sf blockage} } (\theta, \phi) $ for 
the different blockage scenarios. From this plot, we observe a median loss of $3.3$ and $4.5$ dB for the 
$0$ mm air gap, and $2.7$ and $3.5$ dB for the $1$ mm air gap scenarios, respectively. The corresponding 
$90$-th percentile loss values are $8.1$, $11.6$, $7.1$ and $9.4$ dB. Losses decrease by $\approx 1$ 
dB from the $0$ mm to $1$ mm air gap, whereas losses increase by a similar amount for one to two 
fingers. These observations show that the hand can indeed substantially deteriorate the link performance
relative to Freespace operation, requiring careful remediation.

\begin{figure*}[htb!]
\begin{center}
\begin{tabular}{cc}
\includegraphics[height=1.9in,width=2.7in]{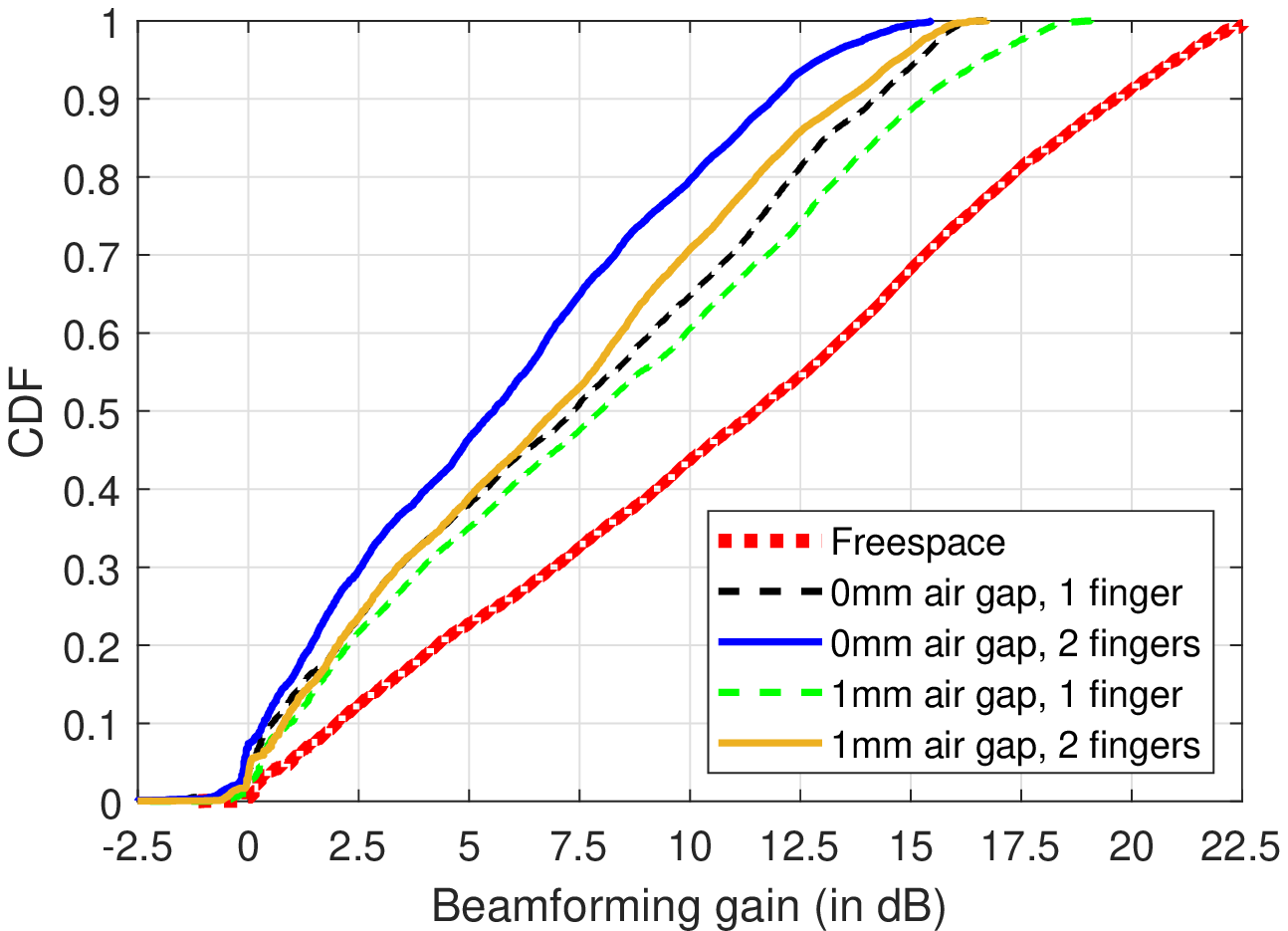}
&
\includegraphics[height=1.9in,width=2.7in]{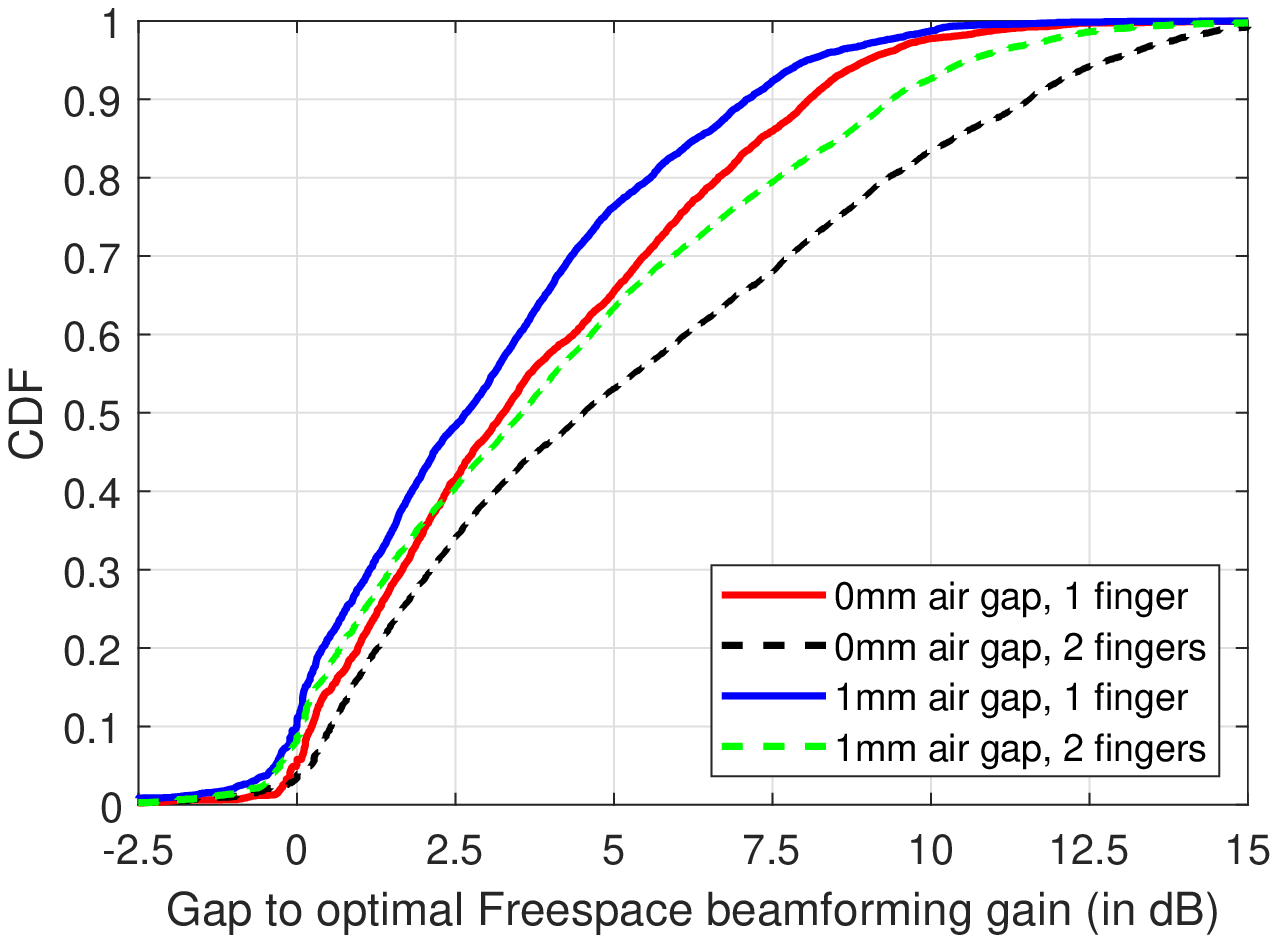}
\\
(a) & (b)
\\
\includegraphics[height=1.9in,width=2.7in]{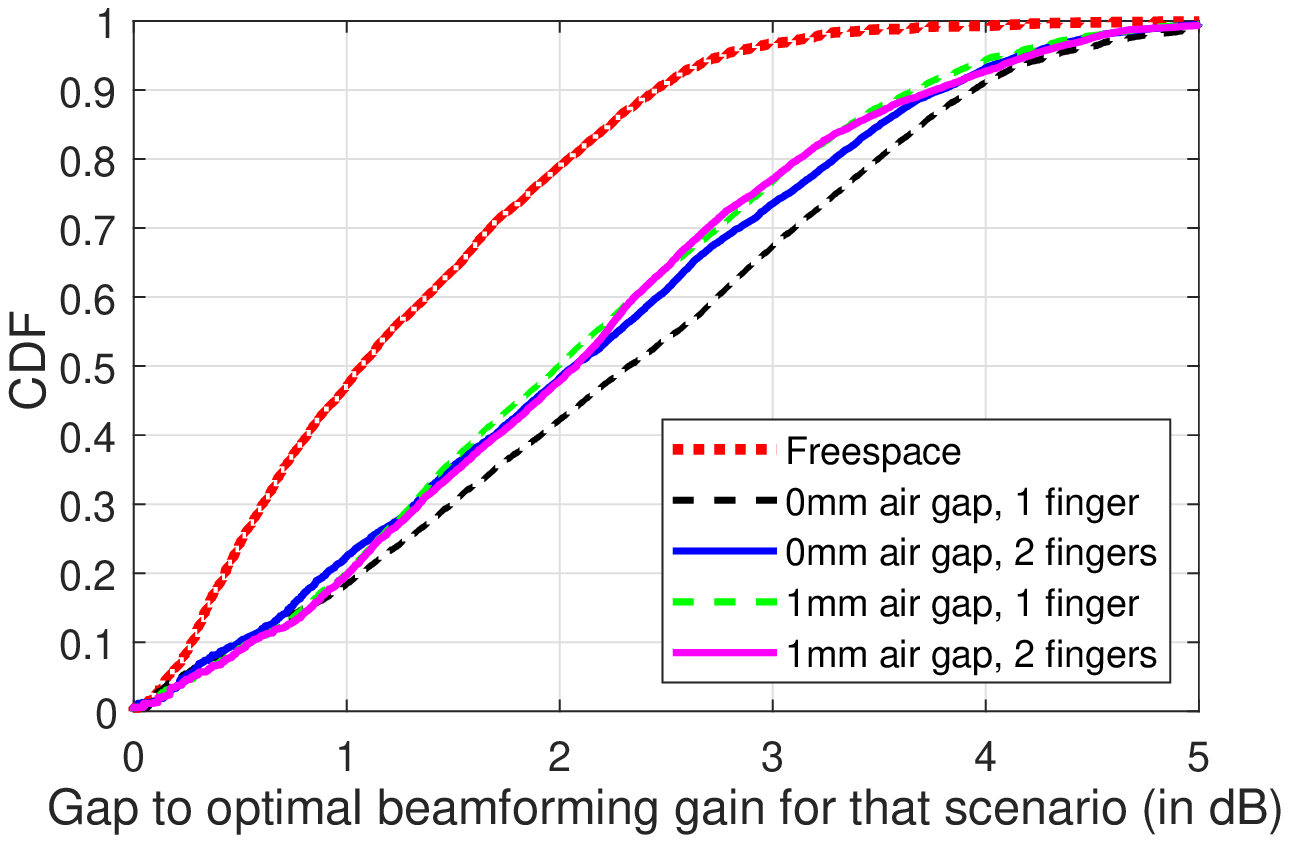}
&
\includegraphics[height=1.9in,width=2.7in]{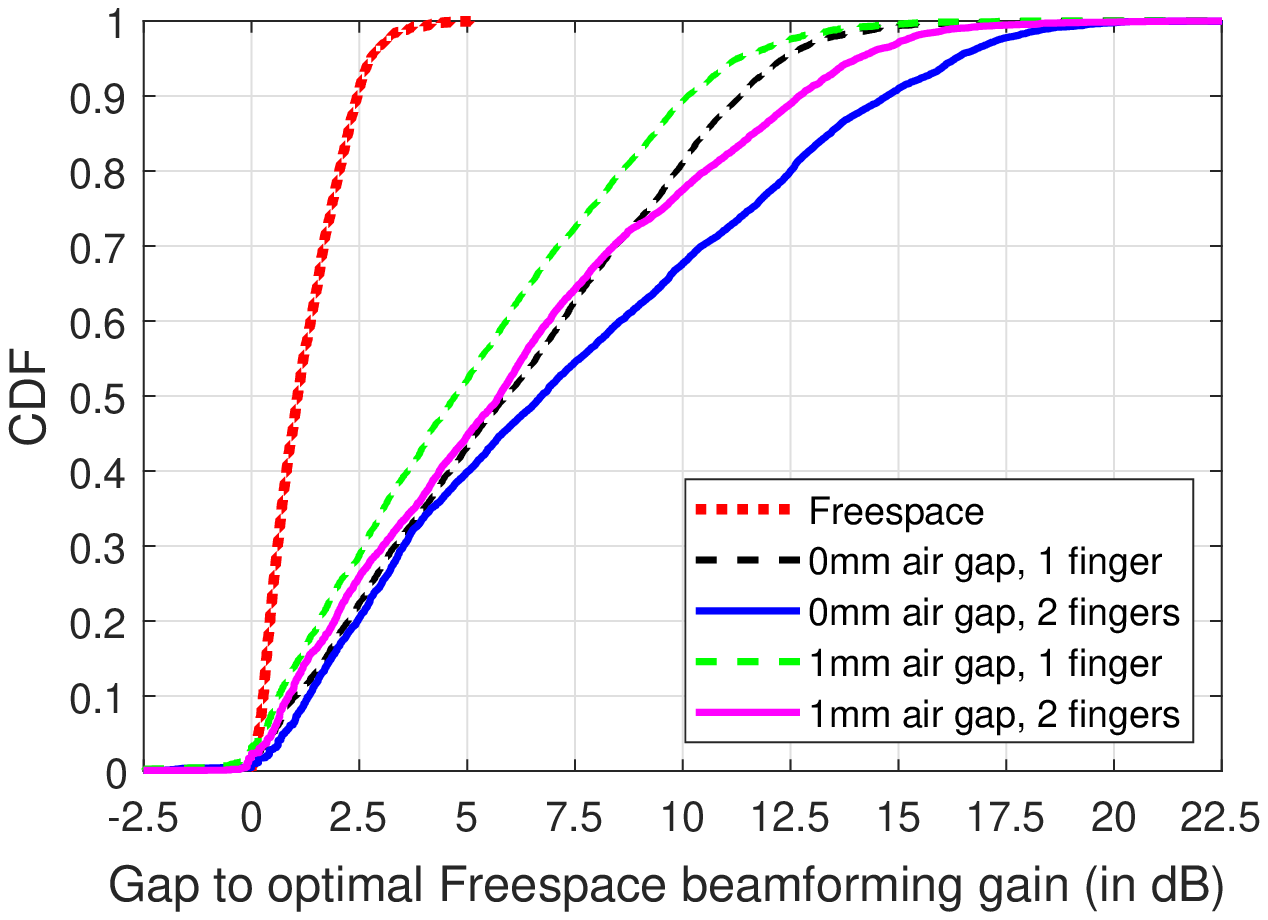}
\\
(c) & (d)
\end{tabular}
\caption{\label{fig_opt}
(a) CDF of optimal beamforming gain under all scenarios.
(b) Loss in optimal beamforming gain under blockage scenarios relative to optimal beamforming
in Freespace.
Loss in codebook-based beamforming gain under Freespace and blockage scenarios relative
to (c) the optimal scheme for that scenario and (d) optimal beamforming in Freespace.
}
\end{center}
\end{figure*}

\subsection{Finite-Sized Beamforming Codebooks}
\label{sec4b}

In a practical deployment setting, beamforming is performed using a directional analog 
beamforming codebook (of size-$J$ where $J$ is chosen appropriately) denoted as 
${\cal C}_{\sf dir}$ corresponding to a static set of beam weights where each beam weight 
steers energy in a fixed {\em a priori} determined direction. Note that this scheme is a 
low-complexity alternative to MRC and is a good scheme for sparse channels such as 
those seen at millimeter wave frequencies~\cite{raghavan_jstsp,vasanth_gcom15,vasanth_tap2018}.
Let ${ w}_{i,j} , \hsppp i = 1, \cdots, N$ denote the unit-norm beam weights for the $j$-th
directional beam ($j  = 1, \cdots, J$) with
\begin{eqnarray}
{\cal C}_{\sf dir} = \Big\{  w_{i,j} , \hsppp i = 1, \cdots, N, \hsppp j = 1, \cdots, J \Big\}.
\nonumber
\end{eqnarray}
The realized gain with this directional codebook is given as
\begin{eqnarray}
{\bf G}_{ {\sf cbk}, \hsppp \bullet} (\theta, \phi) = \max_{j = 1, \cdots, J} 10 \cdot \log_{10} \left(
\left|\sum_{i = 1}^N  w_{i,j}^{\star} E_{ \bullet, \hsppp i}(\theta, \phi) \right|^2
\right)
\nonumber
\end{eqnarray}
along the direction $(\theta, \phi)$.

\ignore{
\begin{figure*}[htb!]
\begin{center}
\begin{tabular}{ccc}
\includegraphics[height=1.9in,width=2.1in]{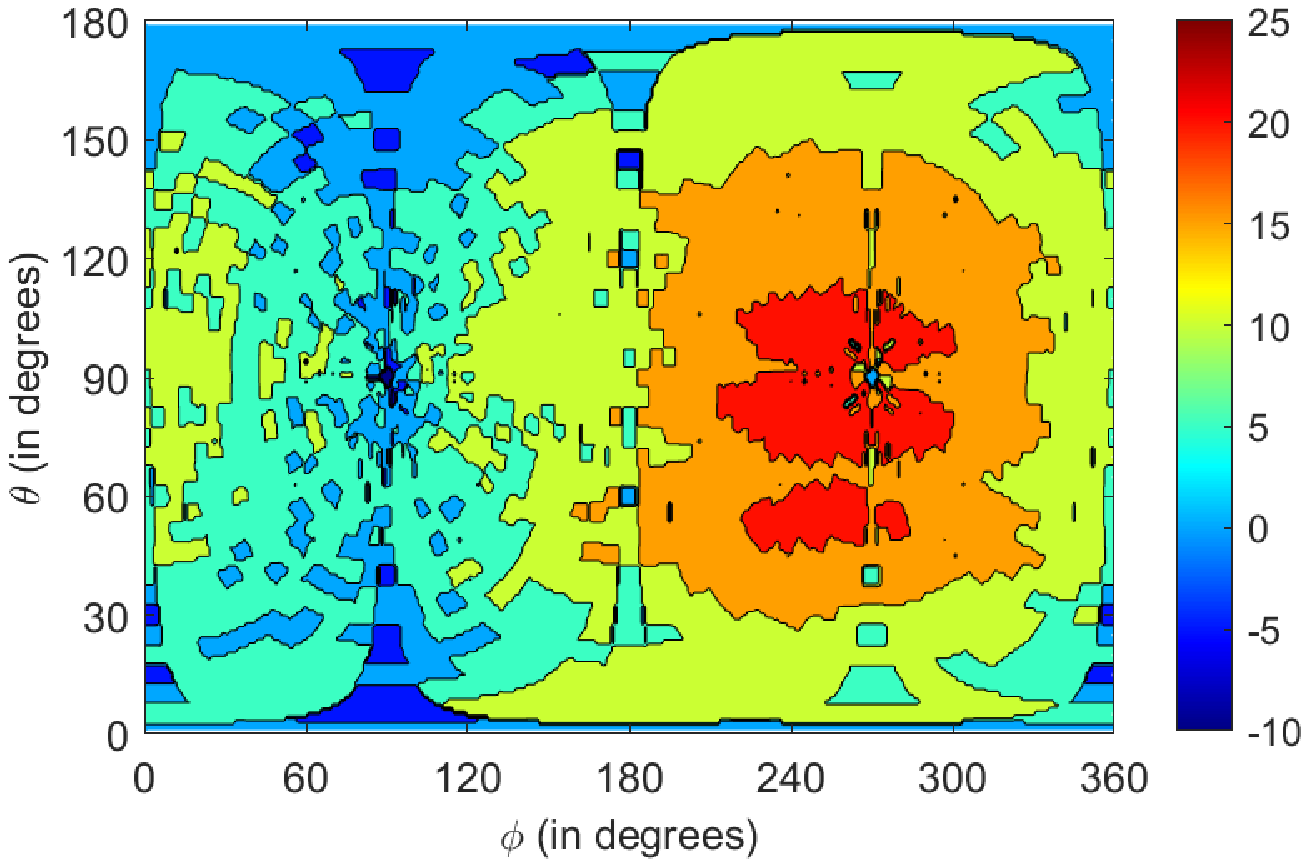}
&
\includegraphics[height=1.9in,width=2.1in]{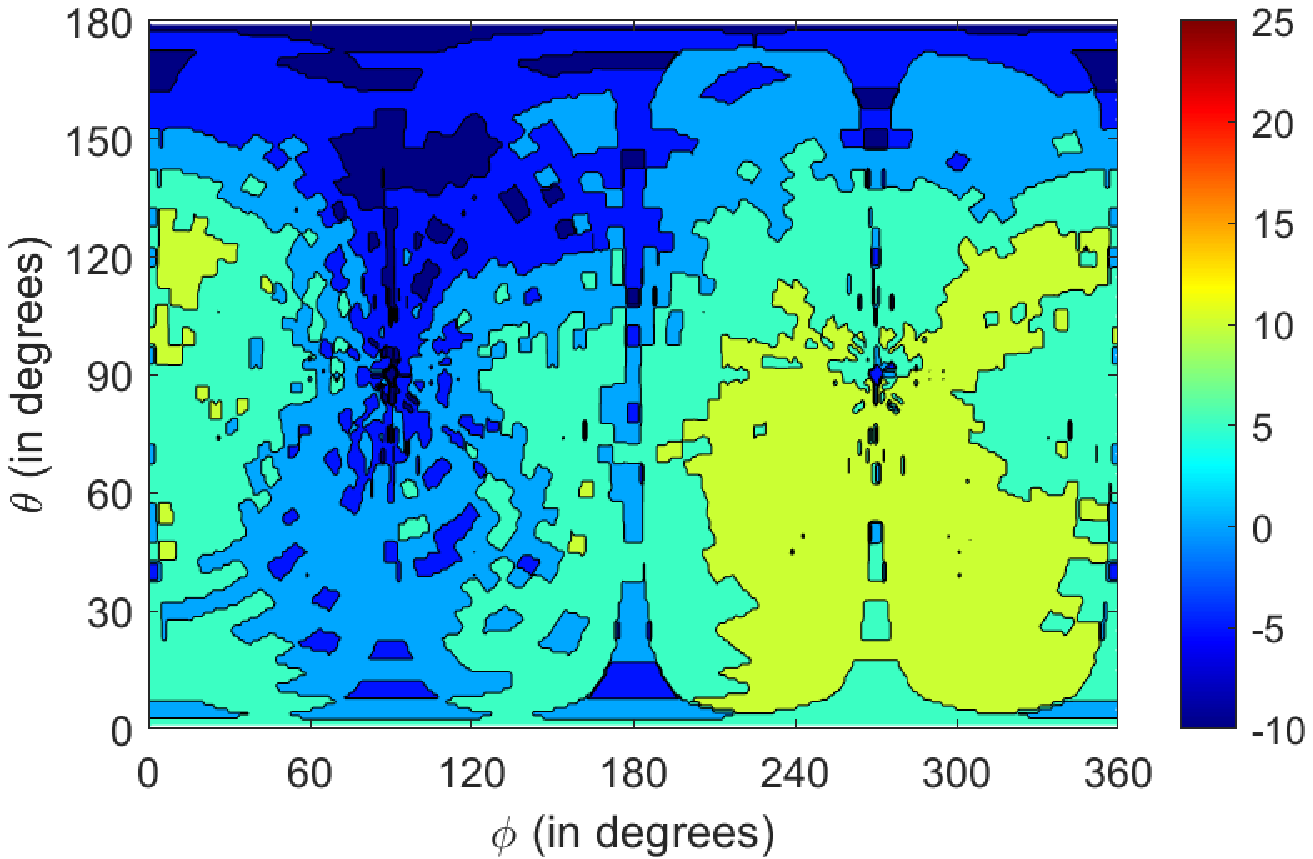}
&
\includegraphics[height=1.9in,width=2.1in]{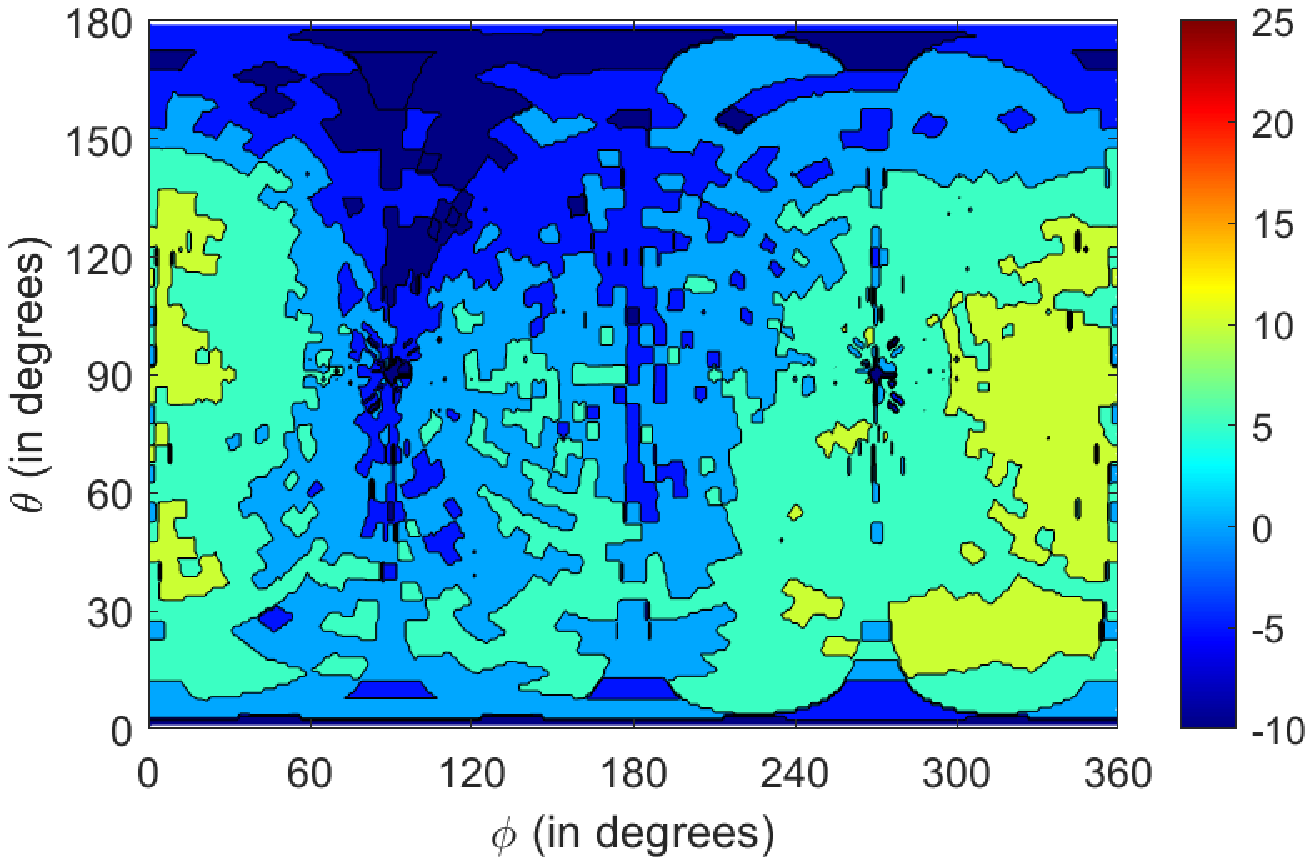}
\\
(a) & (b) & (c)
\end{tabular}
\caption{\label{fig_cbkbf}
Gain with codebook-based beamforming over the sphere in (a) Freespace, and $0$ mm air gap with (b) one and
(c) two fingers, respectively.
}
\end{center}
\end{figure*}
}

It is typical to use $J = N$ for an array of $N$ antenna elements since this arrangement leads to
an $\approx 2.5$ dB cross-over point between adjacent/neighboring beams over the coverage area of the
antenna array. With $J = N = 4$ here, we design beam weights to steer energy in fixed equi-spaced
directions in beamspace. In Figs.~\ref{fig_optbf}(d)-(f), we illustrate the contour plot of the
beamforming gain over the sphere with the codebook-based schemes showing a comparable performance
with the optimal beamforming schemes. In Fig.~\ref{fig_opt}(c), the performance of the scheme with
size-$4$ codebooks is compared with the optimal beamforming scheme in {\em that} scenario. These plots
show that in Freespace, the loss ranges from $1.1$ dB at the median to $2.5$ dB at the $90$-th percentile, 
which is in agreement with the design of a size-$N$ codebook for an $N \times 1$ array. On the
other hand, the median loss in blockage scenarios (with the codebook relative to the optimal scheme)
range from $2$ to $2.3$ dB, whereas the $90$-th percentile ranges from $3.7$ to $3.9$ dB suggesting
that the directional codebook-based beamforming can suffer a significant performance degradation
over the optimal scheme than in Freespace operation.

Also, 
the loss in performance with ${\cal C}_{\sf dir}$ over the optimal performance
in Freespace is plotted in Fig.~\ref{fig_opt}(d). These plots show that the median loss with a $1$ mm
air gap are $4.7$ and $5.8$ dB (one and two fingers) matching up with numbers seen in prior works.
When that $1$ mm air gap becomes $0$ mm, the median numbers increase to $5.9$ and $6.7$ dB --- an
increase of loss by $\approx 1$ dB. The corresponding $90$-th percentile loss numbers are $10.1$, $11.3$,
$12.7$ and $14.7$ dB, which appears to be significant to warrant a dramatic loss in link performance.
These observations motivate the need to improve performance over ${\cal C}_{\sf dir}$, which is the
subject of the next section.

\ignore{
\begin{figure*}[htb!]
\begin{center}
\begin{tabular}{cc}
\includegraphics[height=1.9in,width=2.7in]{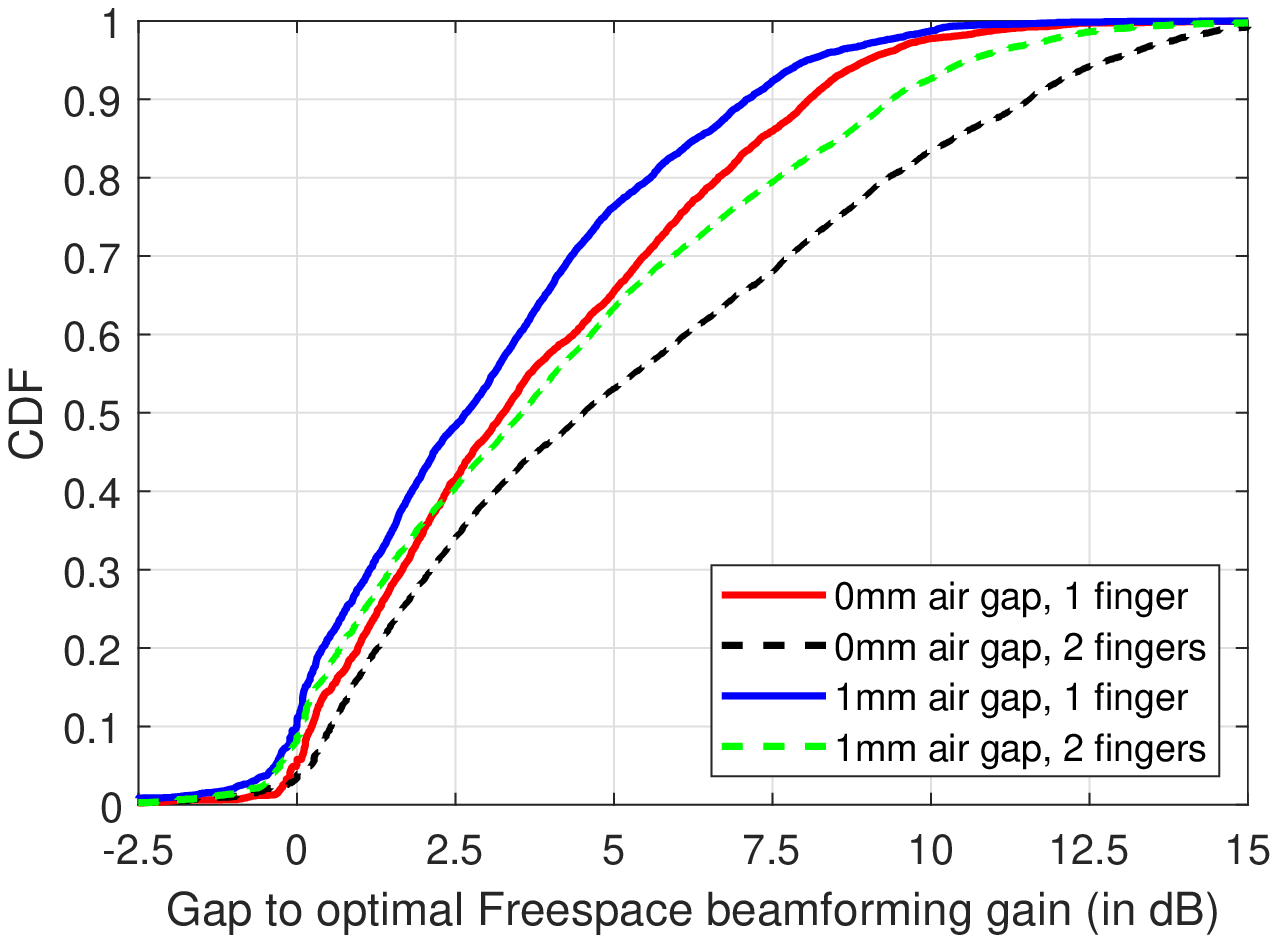}
&
\includegraphics[height=1.9in,width=2.7in]{fig_opt_minus_cbk_0and1mmairgap_v2.eps}
\\
(a) & (b)
\end{tabular}
\caption{\label{fig_opt}
(a) Loss of optimal beamforming under blockage scenarios relative to optimal beamforming in Freespace.
(b) Loss of static codebook-based beamforming relative to optimal beamforming under blockage scenarios.
}
\end{center}
\end{figure*}
}

\section{Blockage Mitigation Via Non-Directional Codebooks}
\label{sec5}

\begin{figure*}[htb!]
\begin{center}
\begin{tabular}{ccc}
\includegraphics[height=1.9in,width=2.1in]{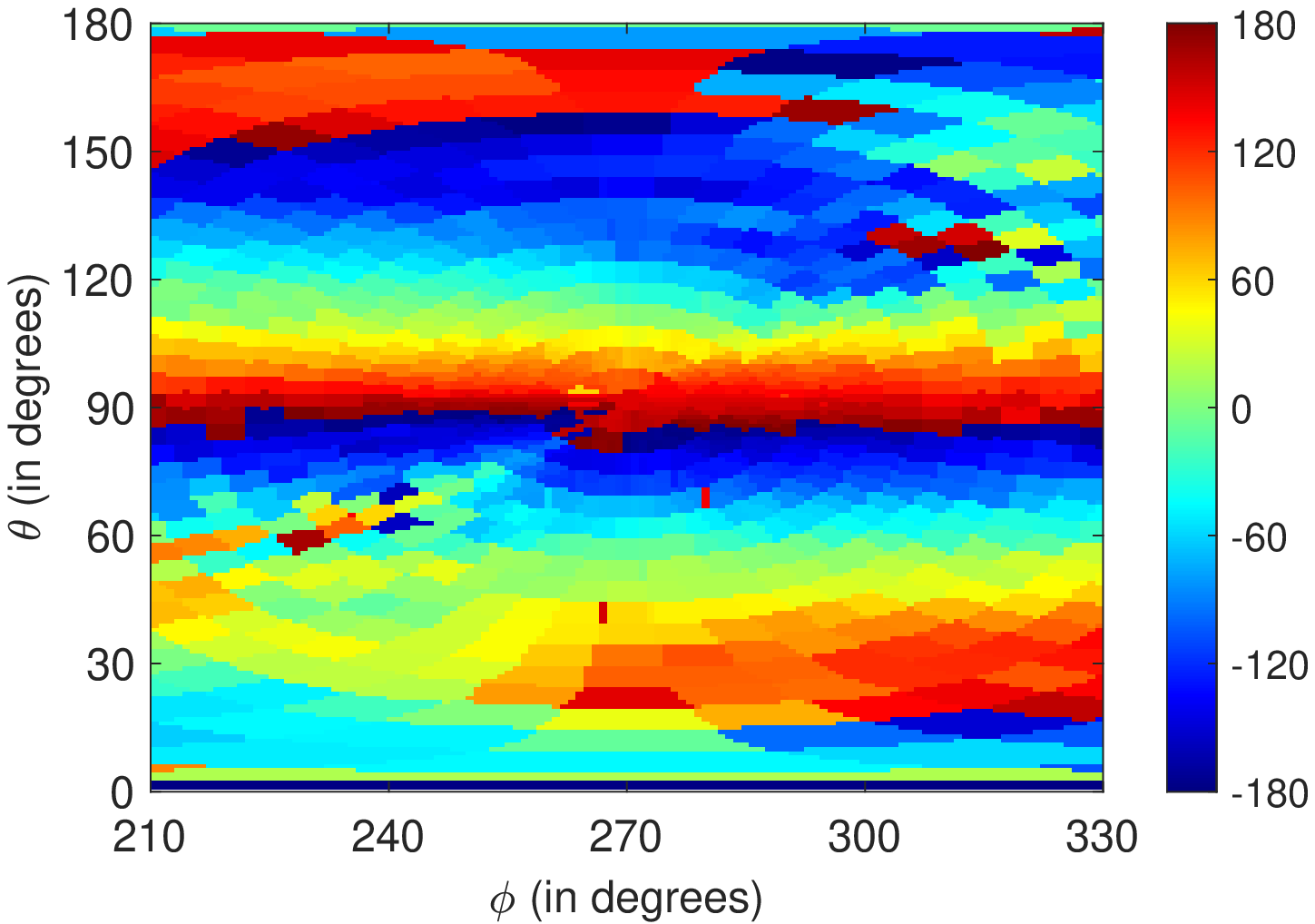}
&
\includegraphics[height=1.9in,width=2.1in]{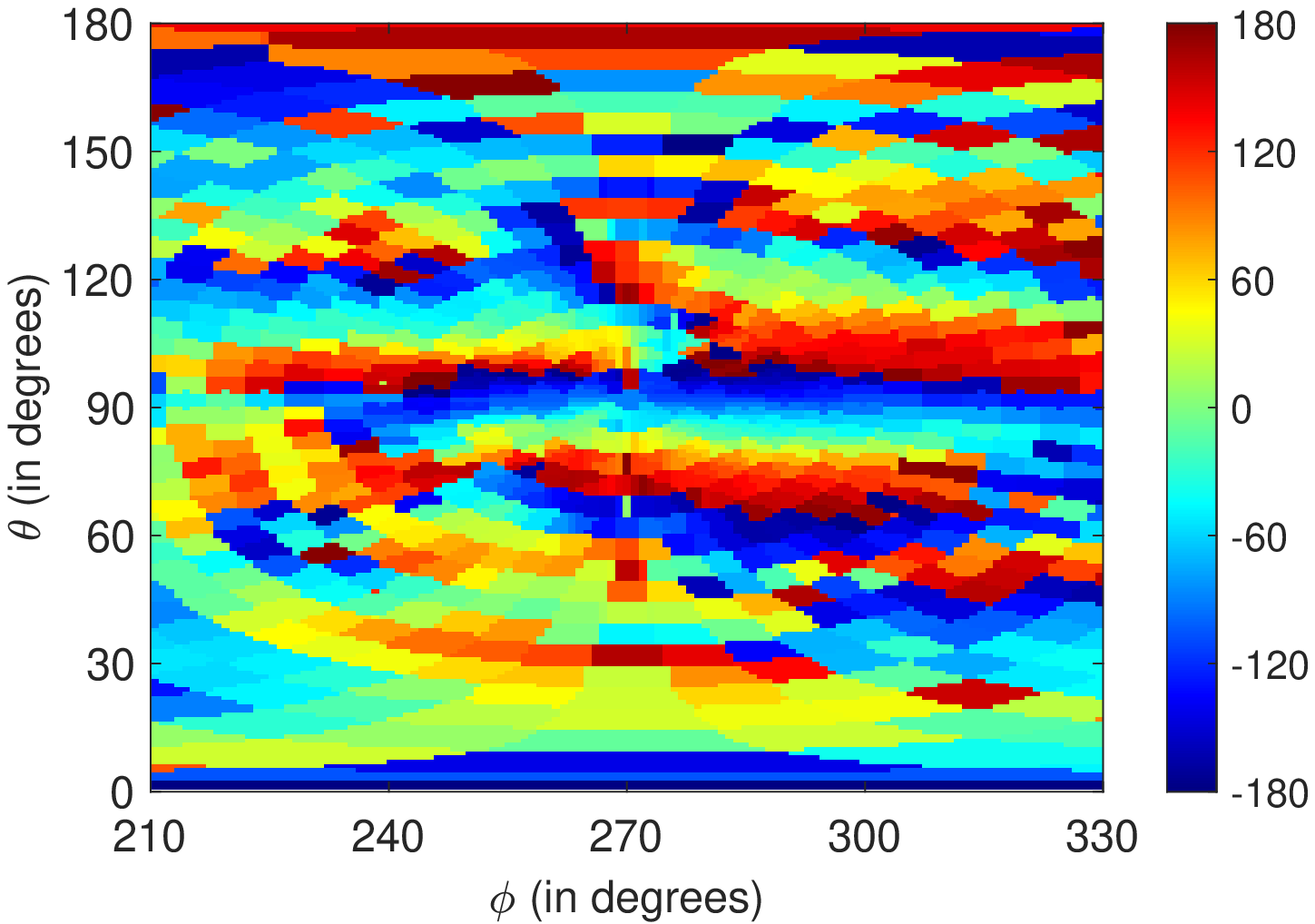}
&
\includegraphics[height=1.9in,width=2.1in]{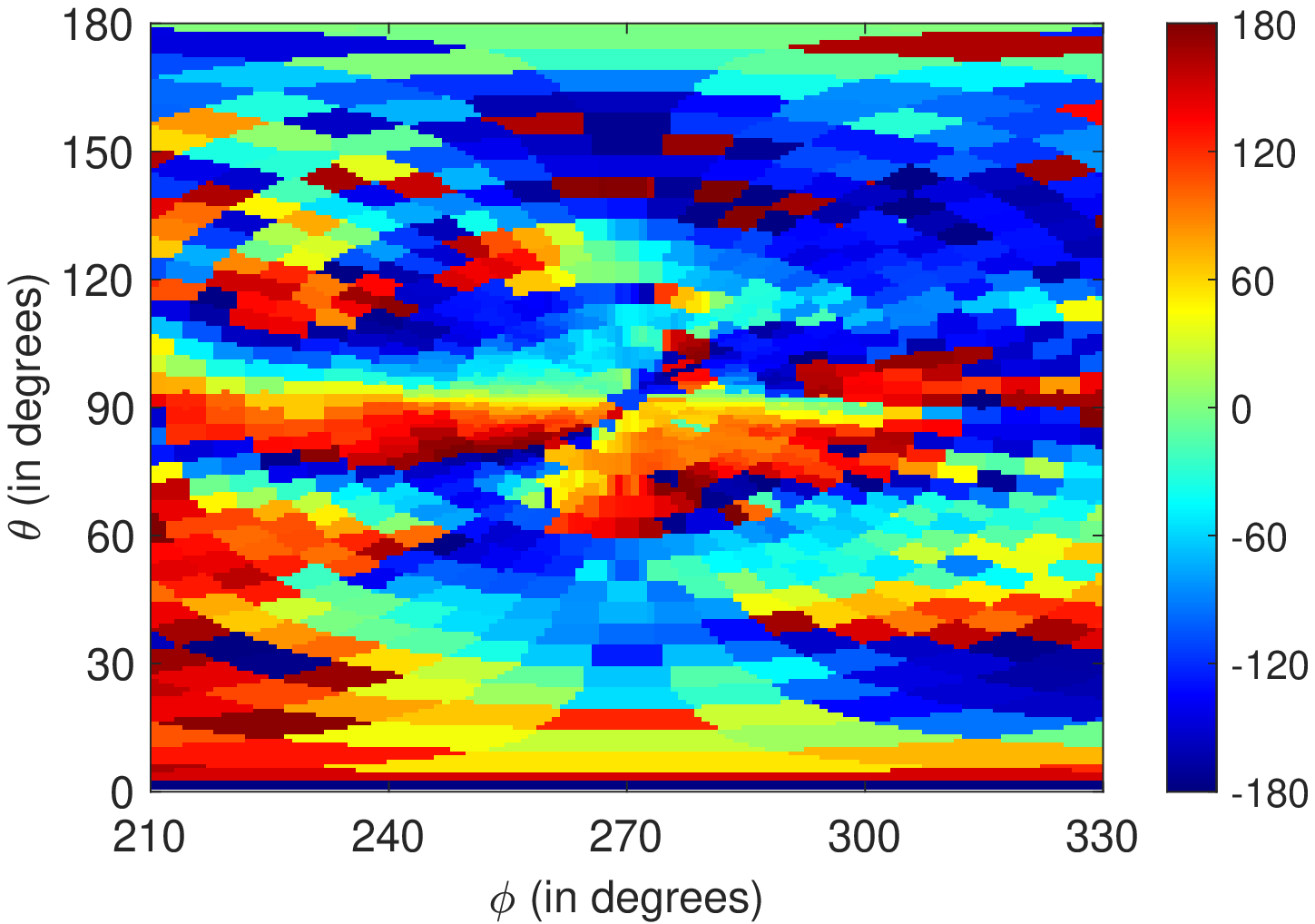}
\\
(a) & (b) & (c)
\end{tabular}
\caption{\label{fig_phase}
Phase response (a) with Freespace measurements and with $0$ mm air gap corresponding to (b) one and (c) two fingers. }
\end{center}
\end{figure*}

\subsection{A Closer Look at the Impact of Blockage on Phases}

Towards improving the performance of ${\cal C}_{\sf dir}$, we start with a more careful study
of the true impact of blockage on the phases seen by the antenna elements. In 
Figs.~\ref{fig_phase}(a)-(c), the phase response (denoted as $\angle{ \Delta E_{ \bullet }(\theta, \phi) }$)
for Antenna 2 with respect to Antenna 1 is plotted for the Freespace case, $0$ mm air gap with one and two fingers,
respectively (where $\bullet \in \{ {\sf free},  {\sf blockage} \}$). Note that 
\begin{eqnarray} 
\angle{ \Delta E_{ \sf free }(\theta, \phi) } & = & \angle{ E_{ {\sf free}, \hsppp 2 }(\theta, \phi) } - 
\angle{ E_{ {\sf free}, \hsppp 1 }(\theta, \phi) } \nonumber \\ 
\angle{ \Delta E_{ \sf blockage }(\theta, \phi) } & = & \angle{ E_{ {\sf blockage}, \hsppp 2 }(\theta, \phi) } -
\angle{ E_{ {\sf blockage}, \hsppp 1 }(\theta, \phi) }.  \nonumber 
\end{eqnarray} 
From a visual perspective, the
presence of the hand phantom leads to significant phase distortions captured as distinct color mixings within
expected regions of contiguous phase behavior (from the Freespace plot in Fig.~\ref{fig_phase}(a)).
To quantitatively capture the extent of this phase mixing, we define the following metric:
\begin{eqnarray}
{\sf Measure} {\hspace{0.04in}} {\sf of} {\hspace{0.04in}} {\sf phase}
{\hspace{0.04in}} {\sf mixing} = {\bf E}_{\theta, \phi}  \Big[
\left| \nabla_{\theta} \left\{ \angle{ \Delta E_{ \bullet }(\theta, \phi) } \right\} \right|
\Big].
\nonumber
\end{eqnarray}
Since the $4 \times 1$ array is placed on the Z axis, the phase response in the ideal scenario varies
only over $\theta$ with no variations over $\phi$ motivating the above choice of the metric, which
captures the average of the absolute directional gradient (direction of interest being $\theta$).
In the practical case where
measurements are made in Freespace, from Fig.~\ref{fig_phase}(a), we observe that there are jumps in phases
over $\theta$ (and significantly smaller jumps over $\phi$), which when averaged would lead to a small value for the phase
mixing metric. On the other hand, if there is a considerable {\em volatility} (or spread) in phases as the hand
phantom can induce, the corresponding values for the phase mixing metric can be large. Thus, a small value
for the phase mixing metric captures a close to ideal phase response and a large value captures a significant
phase distortion.

With this context, we compute the phase mixing metric from the measurement data to be
$13.0^{\sf o}$, $21.2^{\sf o}$ and $20.6^{\sf o}$ for the Freespace, $0$ mm air gap
cases with one and two fingers, respectively. These observations show that a significant phase distortion
is induced by the hand phantom with the fingers leading to deviation in phase response from Freespace behavior.
While Antenna 2's (relative to Antenna 1) phase behavior is presented here, the conclusion seems to be
similar across all the other antenna pairs as well as across polarizations (not presented here due to space
constraints). In general, the presence of the fingers of
the hand on top of/near the antenna module leads to angle- and antenna element-dependent amplitude and phase
distortions due to multiple sets of reflections of the radiation from the antenna element (radiator) by the
indentations of the fingers. These distortions can be modeled as:
\begin{eqnarray}
E_{ {\sf blockage},\hsppp i}(\theta,\phi) =
E_{ {\sf free},\hsppp i}(\theta,\phi) \cdot \underbrace{ A_i(\theta, \phi) e^{j P_i(\theta, \phi)}
}_{\sf Distortion} , \hspp i = 1, \cdots, N
\nonumber
\end{eqnarray}
where $A_i(\theta, \phi)$ and $P_i(\theta, \phi)$ capture the distortions illustrated in
Figs.~\ref{fig_amp} and~\ref{fig_phase}, respectively.


The intuitive explanation behind the poor performance of ${\cal C}_{\sf dir}$ relative
to the optimal scheme in Sec.~\ref{sec4b} is the following. Note that ${\cal C}_{\sf dir}$
consists of beam weights that steer beams towards fixed directions in beamspace. Directional
beams are well understood to produce a low-rank approximation of the dominant eigen-modes of a
sparse channel such as those encountered at millimeter wave
frequencies~\cite{raghavan_jstsp,vasanth_tap2018}. On the other hand, closely approximating the
eigen-modes of the channel (and in particular, the optimal MRC solution)
requires the use of a linear combination of beams steered along different directions and the
use of high-resolution phase shifter and gain controls at each antenna element~\cite{vasanth_gcom15}.
Furthermore, the optimal beamforming weights in the blockage scenario are given as
\begin{eqnarray}
w_i \Big|_{\sf opt} = \frac{  E_{ {\sf free}, \hsppp i}(\theta,\phi) \cdot
A_i(\theta, \phi) e^{j P_i(\theta, \phi)} }{
\sqrt{ \sum_{i = 1}^N |  E_{ {\sf free},\hsppp i}(\theta,\phi) |^2 \cdot
A_i(\theta, \phi)  ^2 }},
\nonumber
\end{eqnarray}
which requires knowledge of the distortion. A static codebook of steered beams (e.g., ${\cal C}_{\sf dir}$)
cannot realize such beam weights leading to its relatively high loss, as illustrated in Fig.~\ref{fig_opt}(c). 
Note that the entries of ${\cal C}_{\sf dir}$ are quantized with a $B = 5$ bit phase shifter (with a 
$11.25^{\sf o}$ quantization spread). Thus, the phase mixing by the hand can lead to destructive 
interference with the entries of ${\cal C}_{\sf dir}$.  
The above observations motivate that any effective blockage mitigation strategy has to address the
amplitude and phase distortions.

Given that the amplitude and phase distortions are a function of
the user's hand properties, hand grip, material property variations over frequency, antenna array
properties (e.g., array size and geometry, impact of housing), etc., mitigating these distortions
by learning them appears difficult. With a lack of ability in learning them, these distortions
essentially appear to be {\em random} from the perspective of beamforming. Thus, we consider a
robust approach and design an enhancement to ${\cal C}_{\sf dir}$ captured by a set of phase shifters
and/or amplitude controls that can sample the space of all possible phases and/or amplitudes with
low overhead. This design is described next. 

\subsection{Proposed Non-Directional Codebook Design}
For this, we start by considering a $B$-bit phase shifter that can (ideally) produce $2^B$ phase possibilities:
\begin{eqnarray}
\phi_k = \frac{2 \pi \cdot k }{2^B} , \hspp k = 0, \cdots, 2^B - 1.
\nonumber
\end{eqnarray}
We then consider a codebook enhancement ${\cal C}_{\sf enh, \hsppp phase}$ of size-$(2^B)^{N-1}$ where
\begin{eqnarray}
{\cal C}_{\sf enh, \hsppp phase} = \Big\{
u_{k_2, \hsppp \cdots, \hsppp k_{N}}, \hspp k_{\ell} = 0, \cdots, 2^B - 1, \hspp
\ell = 2, \cdots , N
\Big\}, 
\nonumber
\end{eqnarray}
with each set of beam weights being of the functional form:
\begin{eqnarray}
u_{k_2, \hsppp \cdots, \hsppp k_{N}}  = \frac{1}{ \sqrt{N}} \cdot
\left[ \begin{array}{c}
1 \\
e^{j \phi_{k_2} } \\
\vdots \\
e^{j \phi_{k_N} }
\end{array}
\right].
\nonumber
\end{eqnarray}
Note that only the relative phases of the antenna elements with respect to the first antenna
matters and thus without loss in generality, we can set the first phase term ($\phi_{k_1}$) to be
$0$ for all the codebook entries. The basic motivation behind the structure of
${\cal C}_{\sf enh, \hsppp phase}$ is to sample each antenna element with a $B$-bit phase
shifter with the best set of beam weights from ${\cal C}_{\sf enh, \hsppp phase}$ being the
closest de-randomizer of the phase distortions induced by the hand. The effective role of the
de-randomizer is to incorporate the impact of the hand distortions in the beam weights used,
thereby matching the beam weights to the effective channel response as well as the hand effects
and thus improving the realized array gains.

Since blockage induces both amplitude and phase distortions, the optimal beam weights for
this scenario need to incorporate a search over {\em both} amplitudes and phases. Unlike phases
with a limited range of $2 \pi$, approximating the amplitude information can lead to a quick
increase in codebook size and therefore the overhead associated with learning these beam weights.
Thus, to overcome this complexity, we consider a beam training procedure with $N$ beams, each
of which excites only one of the $N$ antenna elements at any instant. Let ${\sf S}_i, \hsppp
i = 1, \cdots, N$ denote the estimated signal strength with the $i$-th beam that excites the
$i$-th antenna.
This beam training is performed {\em after} the introduction of hand blockage so that ${\sf S}_i$ can
be estimated with the presence of the hand.

Based on these signal strengths, we consider a codebook enhancement where
\begin{eqnarray}
{\cal C}_{\sf enh, \hsppp phase, \hsppp amp} = \Big\{
v_{k_2, \hsppp \cdots, \hsppp k_{N}}, \hspp k_{\ell} = 0, \cdots, 2^B - 1, \hspp
\ell = 2, \cdots , N
\Big\}, 
\nonumber
\end{eqnarray}
with each set of beam weights being of the functional form:
\begin{eqnarray}
v_{k_2, \hsppp \cdots, \hsppp k_{N}}  = \frac{1}{ \sqrt{  \sum_{i=1}^N {\sf S}_i }} \cdot
\left[ \begin{array}{c}
\sqrt{ {\sf S}_1 } \\
\sqrt{ {\sf S}_2 } \cdot e^{j \phi_{k_2} } \\
\vdots \\
\sqrt{ {\sf S}_N } \cdot e^{j \phi_{k_N} }
\end{array}
\right].
\nonumber
\end{eqnarray}
As before, we can set $\phi_{k_1} = 0$. In the above structure, instead of searching for the
amplitude of the $i$-th antenna element, we approximate it by the normalized square root of the
signal strength based on selecting the $i$-th antenna element. Note that instead of using the 
true/estimated ${\sf S}_i$, if we used
${\sf S}_i = \frac{1}{N}$ for all $i$, then $v_{k_2,  \hsppp \cdots, \hsppp k_{N}}$ reduces to
$u_{k_2, \hsppp \cdots, \hsppp k_{N}}$.

\subsection{Theoretical Performance Comparisons}
We now compare the performance of ${\cal C}_{\sf enh, \hsppp phase, \hsppp amp}$ with
${\cal C}_{\sf enh, \hsppp phase}$ and ${\cal C}_{\sf dir}$ in the blockage setting. For
this, we consider a channel matrix ${\sf H}$ corresponding to $L$ dominant clusters over
which beamformed transmissions are used at both the base station (with $M$ antenna elements)
and UE (with $N$ antenna elements) ends. Let this $N \times M$ channel matrix ${\sf H}$ be given 
as~\cite{saleh}
\begin{eqnarray}
{\sf H} = \sum_{\ell = 1}^L \alpha_{\ell} \cdot {\bf E}_{\sf blockage}(\theta_{ {\sf R}, \hsppp \ell},
\phi_{ {\sf R}, \hsppp \ell} ) {\bf a}_{\sf T} (\theta_{ {\sf T}, \hsppp \ell},
\phi_{ {\sf T}, \hsppp \ell} )^H
\label{eq_chan}
\end{eqnarray}
where $\alpha_{\ell}$, $\theta_{ {\sf R}, \hsppp \ell}$, $\phi_{ {\sf R}, \hsppp \ell}$,
$\theta_{ {\sf T}, \hsppp \ell}$ and $\phi_{ {\sf T}, \hsppp \ell}$ denote the complex gain,
elevation and azimuth angles at the UE and base station ends, respectively, and $(\cdot)^H$ 
denotes the complex conjugate Hermitian operation. The $N \times 1$
vector ${\bf E}_{\sf blockage}(\theta, \phi)$ captures the electric field vector
at the UE end under blockage setting and is given as
\begin{eqnarray}
{\bf E}_{\sf blockage}(\theta, \phi) = \left[ \begin{array}{c}
E_{ {\sf blockage}, \hsppp 1 }(\theta, \phi) \\
\vdots \\
E_{ {\sf blockage}, \hsppp N}(\theta, \phi)
\end{array}
\right],
\nonumber
\end{eqnarray}
with 
the $M \times 1$ vector ${\bf a}_{\sf T}(\theta, \phi)$ capturing the array steering vector at the 
base station side over the $(\theta, \phi)$ angle pair. Note that the model considered in~(\ref{eq_chan})
is the same as the Saleh-Valenzuela model~\cite{saleh}, popularly used in studies of
millimeter wave systems, with the difference being that the array steering vector at the UE end
is replaced with the electric field vector to capture the impact of the UE housing and material
properties and polarization mismatches/impairments on the steering vector.

We assume that the base station and the UE beamform along unit-norm vectors ${\bf f}$ (of size
$M \times 1$) and ${\bf g}$ (of size $N \times 1$) to lead to the following scalar 
input-output equation model: 
\begin{eqnarray} 
\widehat{\sf s} = \sqrt{\rho} \cdot {\bf g}^H {\sf H} {\bf f} {\sf s} + {\bf g}^H {\bf n} 
\nonumber 
\end{eqnarray} 
with ${\sf s}$ being the scalar input from a certain constellation, $\widehat{\sf s}$ being its estimate, 
${\bf n} \in {\cal CN}({\bf 0}, {\bf I}_N)$ being the additive white Gaussian noise and $\rho$ 
being the transmit power. With this setup, the received ${\sf SNR}$ is given as
\begin{eqnarray}
{\sf SNR}_{\sf rx} = \frac{ \rho \cdot |{\bf g}^H {\sf H} {\bf f}|^2 } 
{ {\bf E} \left[ |{\bf g}^H {\bf n}|^2 \right] } = 
\frac{ \rho \cdot |{\bf g}^H {\sf H} {\bf f}|^2 } 
{ {\bf g}^H \cdot {\bf I}_N \cdot {\bf g} } = 
\rho \cdot |{\bf g}^H {\sf H} {\bf f}|^2.
\nonumber
\end{eqnarray}
Without loss in generality, we assume that $\rho = 1$ and $|\alpha_1| \geq \cdots
\geq |\alpha_L|$. Thus, the angles corresponding to the dominant cluster are $\theta_{ {\sf R}, \hsppp 1}$,
$\phi_{ {\sf R}, \hsppp 1}$, $\theta_{ {\sf T}, \hsppp 1}$ and $\phi_{ {\sf T}, \hsppp 1}$. 
We make the practical assumption of the use of a large antenna array at the base station end with
beamforming over a narrow beamwidth to the dominant cluster in the channel. That is,
\begin{eqnarray}
|{\bf a}_{\sf T} (\theta_{ {\sf T}, \hsppp 1},
\phi_{ {\sf T}, \hsppp 1} )^H {\bf f}| & \approx & 1, \hspp {\sf and}
\nonumber \\
{\bf a}_{\sf T} (\theta_{ {\sf T}, \hsppp \ell},
\phi_{ {\sf T}, \hsppp \ell} )^H {\bf f} & \approx & 0, \hspp \ell = 2 , \cdots, L.
\nonumber
\end{eqnarray}
With these assumptions, we have the following simplification:
\begin{eqnarray}
{\sf SNR}_{\sf rx} \approx
|\alpha_1|^2 \cdot \left| {\bf g}^H {\bf E}_{\sf blockage}(\theta_{ {\sf R}, \hsppp 1},
\phi_{ {\sf R}, \hsppp 1} ) \right|^2 \triangleq \widetilde{\sf SNR}_{\sf rx}.
\nonumber
\end{eqnarray}

We first discuss the performance realized with ${\cal C}_{\sf dir}$ under blockage. For this,
we note that
\begin{eqnarray}
\frac{ \max \limits_{ {\bf g} \hsppp \in
\hsppp {\cal C}_{\sf dir} } \widetilde{\sf SNR}_{\sf rx} }
{|\alpha_1|^2}
= \frac{1}{N} \cdot
\max_j \left| \sum_i
| E_ { {\sf free}, \hsppp i} (\theta_{ {\sf R}, \hsppp 1},
\phi_{ {\sf R}, \hsppp 1} ) | \cdot A_i(\theta_{ {\sf R}, \hsppp 1},
\phi_{ {\sf R}, \hsppp 1} ) \cdot e^{j \psi_i   } \right|^2
\nonumber
\end{eqnarray}
where $w_{ij} = \frac{1}{ \sqrt{N}} \cdot e^{j \angle{w_{ij}}}$ and we have defined
\begin{eqnarray}
\psi_i = \angle{ E_{ {\sf free}, \hsppp i } (\theta_{ {\sf R}, \hsppp 1},
\phi_{ {\sf R}, \hsppp 1} ) } + P_i(\theta_{ {\sf R}, \hsppp 1},
\phi_{ {\sf R}, \hsppp 1} ) - \angle{ w_{ij} } , \hspp i = 1, \cdots, N.
\nonumber
\end{eqnarray}
While the performance seen with ${\cal C}_{\sf dir}$ is a function of how blockage impacts
the antenna response, it is important to note that $\angle{w_{ij}}$ is determined
based on beam steering requirements in Freespace (alone). Thus, the $\psi_i$'s have no
constraints on their ranges and $\psi_i \in [0, 2 \pi)$ in general. As a result, some hand holdings
can lead to constructive addition of the antenna responses, whereas some hand holdings can
lead to destructive addition of the antenna responses. In this sense, ${\cal C}_{\sf dir}$
does not lead to a robust performance with blockage with the worst-case blockage performance
(in terms of the phases $\{P_i\}$) being given as
\begin{eqnarray}
\min \limits_{ \{ P_i \} }
\frac{ \max \limits_{ {\bf g} \hsppp \in
\hsppp {\cal C}_{\sf dir} } \widetilde{\sf SNR}_{\sf rx} }
{|\alpha_1|^2}
& = & \min \limits_{ \{ P_i \} }
\frac{1}{N} \cdot
\max_j \left| \sum_i
| E_ { {\sf free}, \hsppp i} (\theta_{ {\sf R}, \hsppp 1},
\phi_{ {\sf R}, \hsppp 1} ) | \cdot A_i(\theta_{ {\sf R}, \hsppp 1},
\phi_{ {\sf R}, \hsppp 1} ) \cdot e^{j \psi_i   } \right|^2
\nonumber  \\
& = &
\min \limits_{a_i \hsppp \in \hsppp \pm 1} \frac{1}{N} \cdot
\max_j \left| \sum_i
| E_ { {\sf free}, \hsppp i} (\theta_{ {\sf R}, \hsppp 1},
\phi_{ {\sf R}, \hsppp 1} ) | \cdot A_i(\theta_{ {\sf R}, \hsppp 1},
\phi_{ {\sf R}, \hsppp 1} ) \cdot a_i \right|^2.
\nonumber
\end{eqnarray}
While the precise choice of $\{a_i\}$ that minimizes the received ${\sf SNR}$ is a function of the
relative antenna strengths seen with blockage across the antenna array, when multiple antenna
elements see comparable signal strengths with blockage, if their relative phases $\{ P_i\}$ are not
aligned up perfectly in relation to $\angle{w_{ij}}$, destructive interference can lead to poor
performance with ${\cal C}_{\sf dir}$. In this context, by choosing the phases $\{ \phi_{k_i} \}$
for ${\cal C}_{\sf enh, \hsppp phase, \hsppp amp}$ and ${\cal C}_{\sf enh, \hsppp phase}$
appropriately, the range of $\psi_i$ can be restricted (or reduced) and received ${\sf SNR}$ performance
can be improved over that of ${\cal C}_{\sf dir}$. Of these two codebook choices, based on
classical tradeoffs between equal gain combining and MRC solutions, it is intuitively
expected that amplitude and phase control can improve performance over phase-only control~\cite{david_egc}.
In particular, let the improvement in received ${\sf SNR}$ using
${\cal C}_{\sf enh, \hsppp phase, \hsppp amp}$ over ${\cal C}_{\sf enh, \hsppp phase}$ be defined as
\begin{eqnarray}
\widetilde{\Delta} {\sf SNR}_{\sf rx} \triangleq
\frac{ \Delta {\sf SNR}_{\sf rx} }{ |\alpha_1|^2 }
=
\frac{
\max \limits_{ {\bf g} \hsppp \in \hsppp {\cal C}_{\sf enh, \hsppp phase, \hsppp amp}  }
\widetilde{\sf SNR}_{\sf rx}
- \max \limits_{ {\bf g} \hsppp \in \hsppp {\cal C}_{\sf enh, \hsppp phase} } \widetilde{\sf SNR}_{\sf rx}  }
{|\alpha_1|^2}.
\nonumber
\end{eqnarray}
We now quantify $\widetilde{\Delta} {\sf SNR}_{\sf rx}$ in Theorem~\ref{thm1}.

\begin{thm}
\label{thm1}
In the high-SNR setting and assuming $B \geq 1$, $\widetilde{\Delta} {\sf SNR}_{\sf rx}$ is bounded as
\begin{eqnarray}
\widetilde{\Delta} {\sf SNR}_{\sf rx} & \geq &
N \cdot {\sf Var}_{\sf blockage} \cdot
\cos^2 \left( \frac{ \pi}{2^B} \right)
- \frac{ 2 \sin^2 \left( \frac{ \pi}{2^B} \right) }{N} \cdot
\left( \sum |E_{{\sf blockage},\hsppp i} (\theta_{ {\sf R}, \hsppp 1},
\phi_{ {\sf R}, \hsppp 1} ) | \right)^2
\label{eq_thm1} \\
& \triangleq & \Delta {\sf SNR}_{\sf rx, \hsppp LB}
\nonumber 
\end{eqnarray} 
where ${\sf Var}_{\sf blockage}$ denotes the variance of the electric field vector under blockage and 
is given as 
\begin{eqnarray} 
{\sf Var}_{\sf blockage} & = &
\frac{ \sum_i |E_{ {\sf blockage},\hsppp i} (\theta_{ {\sf R}, \hsppp 1},
\phi_{ {\sf R}, \hsppp 1} )|^2 }{N} -
\left( \frac{ \sum_i | E_{ {\sf blockage}, \hsppp i}(\theta_{ {\sf R}, \hsppp 1},
\phi_{ {\sf R}, \hsppp 1} ) | }{N} \right)^2.
\nonumber
\end{eqnarray}
\end{thm}
{\vspace{0.1in}}
\begin{proof}
See Appendix~\ref{app_thm1}.
\end{proof}

There are a number of parameters that impact $\Delta {\sf SNR}_{\sf rx, \hsppp LB}$ as defined 
in Theorem~\ref{thm1}. Clearly, $\Delta {\sf SNR}_{\sf rx, \hsppp LB}$ is maximized as $\cos^2 \left( \frac{\pi}
{2^B} \right)$ approaches $1$ (or as $B$ increases). Further, for a choice of $B$ and 
$\{ E_{ {\sf blockage}, \hsppp i}(\theta_{ {\sf R}, \hsppp 1},
\phi_{ {\sf R}, \hsppp 1} ) \}$ such that $\sum_i | E_{ {\sf blockage}, \hsppp i}(\theta_{ {\sf R}, \hsppp 1},
\phi_{ {\sf R}, \hsppp 1} )|$ is held constant, $\Delta {\sf SNR}_{\sf rx, \hsppp LB}$ is maximized
when ${\sf Var}_{\sf blockage}$ reaches its largest value. Note that ${\sf Var}_{\sf blockage}$ reaches its smallest
and largest values when $| E_{ {\sf blockage}, \hsppp i}(\theta_{ {\sf R}, \hsppp 1},
\phi_{ {\sf R}, \hsppp 1} )|$ are equal for all $i$ and only one of the $| E_{ {\sf blockage}, \hsppp i}(\theta_{ {\sf R}, \hsppp 1},
\phi_{ {\sf R}, \hsppp 1} )|$'s dominate all the other, respectively. In other words, $\Delta {\sf SNR}_{\sf rx, \hsppp LB}$
is the largest when the amplitudes seen with blockage lead to the widest disparity across antenna elements and
in this setting, we have
\begin{eqnarray}
\Delta {\sf SNR}_{\sf rx, \hsppp LB} =
\max_i |E_{{\sf blockage},\hsppp i} (\theta_{ {\sf R}, \hsppp 1},
\phi_{ {\sf R}, \hsppp 1} ) |^2 \cdot \left[
\cos^2 \left( \frac{ \pi}{2^B} \right) \cdot \left(1 - \frac{1}{N} \right)
- \frac{ 2}{N} \sin^2 \left( \frac{ \pi}{2^B} \right) \right].
\nonumber
\end{eqnarray}
The above intuition is not surprising (in hindsight) since ${\cal C}_{\sf enh, \hsppp phase, \hsppp amp}$ performs 
amplitude control over ${\cal C}_{\sf enh, \hsppp phase}$, and amplitude control in MRC is necessitated and is useful 
when the amplitude response across antenna elements affects the antenna elements differently. Thus, when the 
hand distorts the effective response across the antenna elements in an unequal manner, the efficacy of 
${\cal C}_{\sf enh, \hsppp phase, \hsppp amp}$ over ${\cal C}_{\sf enh, \hsppp phase}$ or ${\cal C}_{\sf dir}$ 
is amplified, with more unequal the array response, the better the efficacy.

\begin{figure*}[htb!]
\begin{center}
\begin{tabular}{cc}
\includegraphics[height=1.9in,width=2.9in]{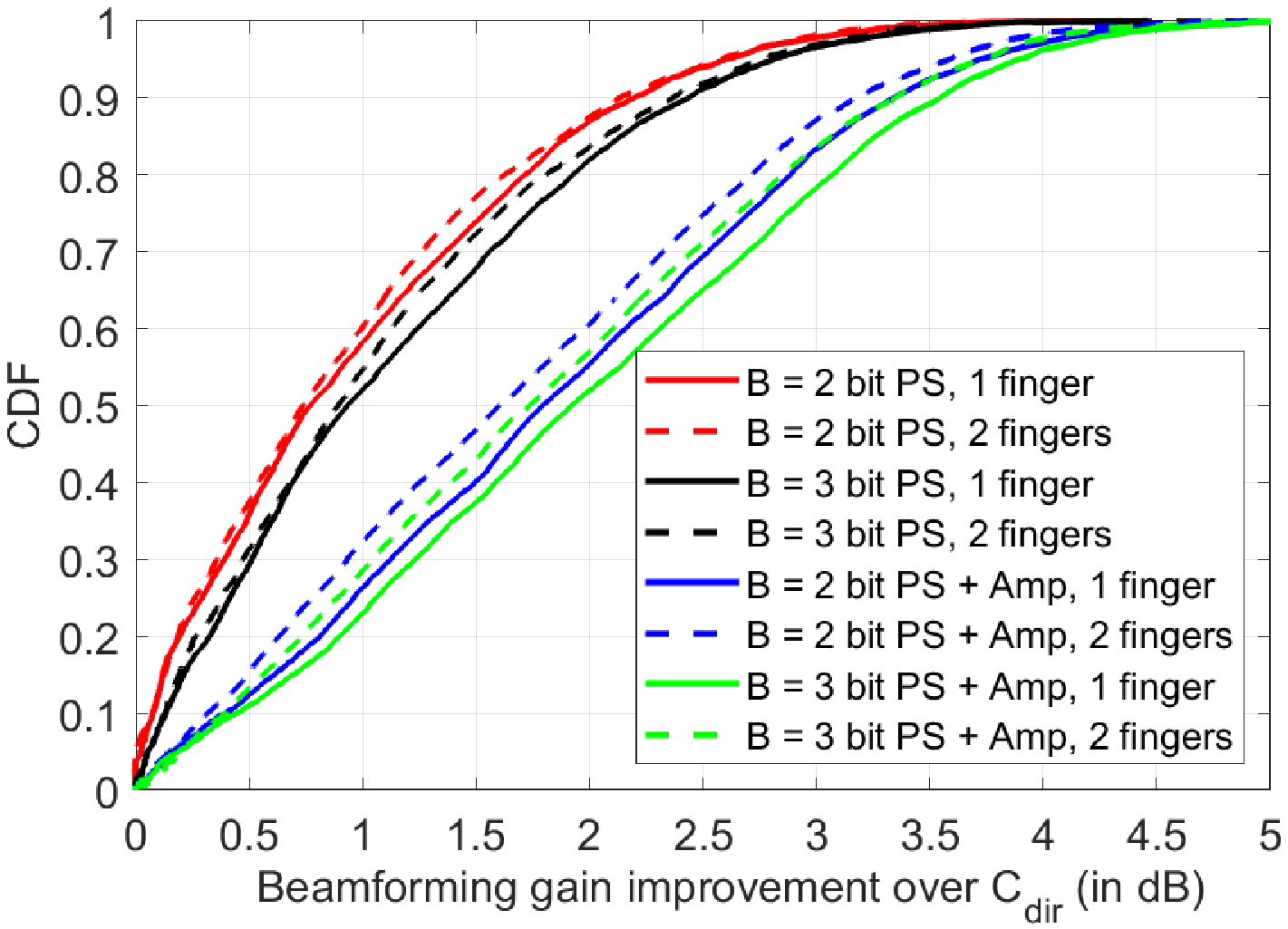}
&
\includegraphics[height=1.9in,width=2.9in]{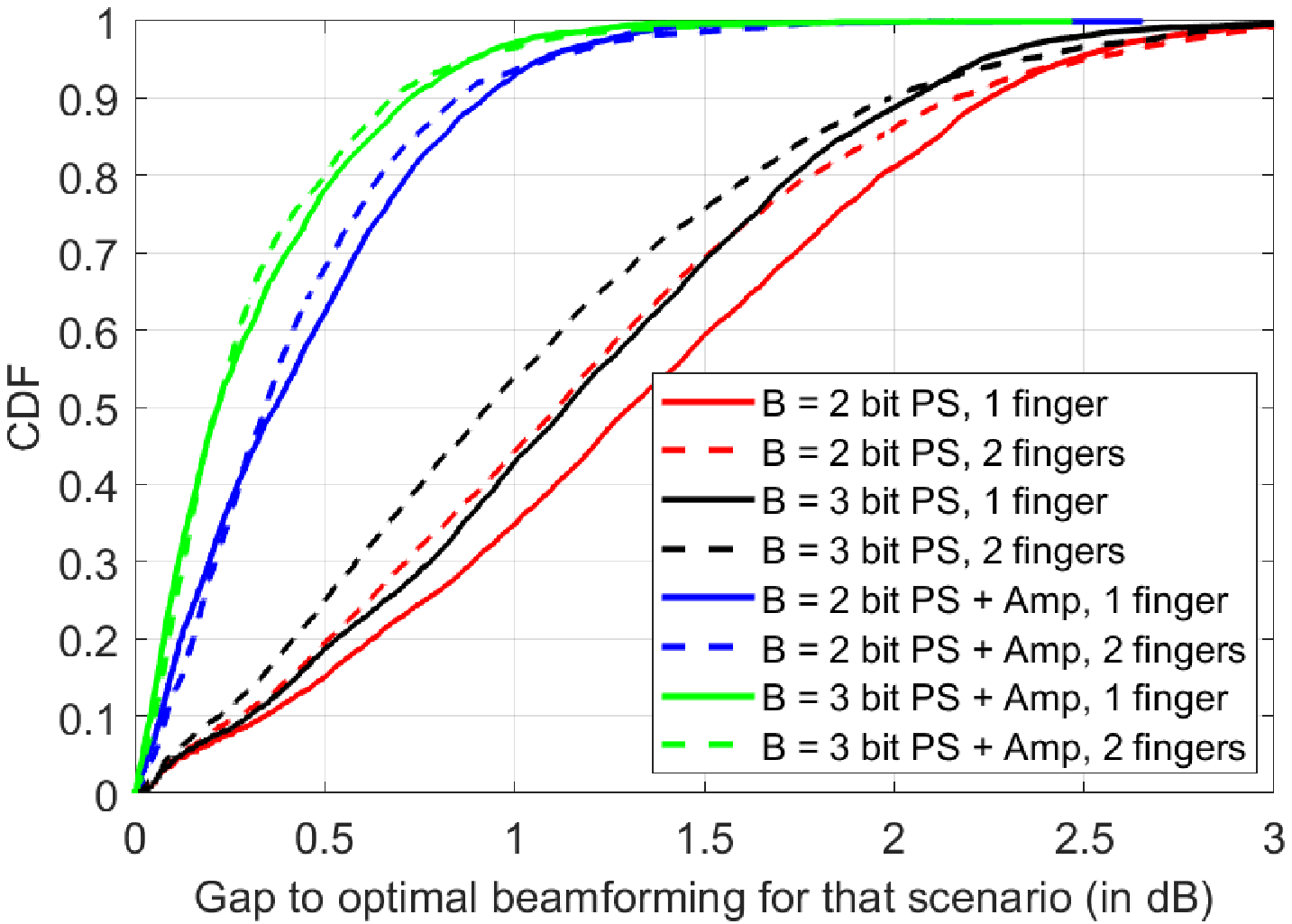}
\\
(a) & (b)
\end{tabular}
\caption{\label{fig_cenh}
(a) Performance improvement over ${\cal C}_{\sf dir}$ for different codebook enhancement schemes.
(b) Gap to optimal beamforming with proposed codebook enhancement schemes.
}
\end{center}
\end{figure*}

\subsection{Numerical Studies} 
Depending on the angle of arrival of the dominant cluster at the UE end, Fig.~\ref{fig_amp}
shows that the amplitude response across antenna elements can be either comparable or not
comparable leading to good performance with ${\cal C}_{\sf enh, \hsppp phase}$ and no need for
${\cal C}_{\sf enh, \hsppp phase, \hsppp amp}$, or the need for
${\cal C}_{\sf enh, \hsppp phase, \hsppp amp}$ to improve blockage performance. To quantify
the performance of de-randomizing the phases and/or amplitudes, we consider four schemes for
numerical evaluation in the $0$ mm air gap case with one and two fingers. For the
first scheme, with $B = 2$, note that the phases of
each antenna element are of the form $\{1, j, -1, -j \}$ and with $N = 4$, we consider a
${\cal C}_{\sf enh, \hsppp phase}$ of size-$64$ ($= (2^B)^{N-1} = (2^2)^3$). For the second scheme, in the
$B = 3$ case, the size of ${\cal C}_{\sf enh,\hsppp phase}$ is $512$ ($= (2^3)^3$). In addition
to these two phase shifter-{\em only} selection schemes, we also consider the amplitude and
phase shifter selection scheme with $B = 2$ and $B = 3$ as the third and fourth schemes, 
respectively.

For these four schemes, Fig.~\ref{fig_cenh}(a) 
plots the beamforming gain improvement with the codebook enhancements over ${\cal C}_{\sf dir}$
for the $0$ mm air gap case with one and two fingers. From these plots, we observe the median,
$80$-th and $90$-th percentile performance improvement of $0.7$, $1.7$ and $2.1$ dB for the first scheme
(substantial numbers in practical settings) suggesting that the fingers of the hand do actually
randomize the phases of different antenna elements which ${\cal C}_{\sf enh, \hsppp phase}$
can de-randomize. Increasing $B$ in the phase shifter selection approach only leads to a marginal
performance improvement (comparable improvement of $0.9$, $1.9$ and $2.4$ dB) suggesting that most 
of the gains with phase shifter selection are
captured with the $B = 2$ bit phase shifter choice. On the other hand, addition of the signal
strength to mirror an MRC-type solution can lead to significant gains ($1.6$ dB at the median
and $3.2$ dB at the $90$-th percentile). Similar numbers for $B = 3$ phase and amplitude control 
over phase-only control 
are $1.7$ dB gain at median and $3.3$ dB at the $90$-th percentile,
again reinforcing that $B = 2$ is sufficient. Thus, it is important to consider a hand blockage 
mitigation strategy that mirrors and accounts for the signal strength and phase variations 
seen across the antenna array commensurate with the hand position. 

To complement this study, Fig.~\ref{fig_cenh}(b)
shows the gap in performance between these schemes and the optimal MRC scheme (which can be
viewed as the ``unrecovered gain''). The median, $80$-th and $90$-th percentile values of unrecovered
gains with the $B = 2$ bit phase shifter based search are $1.1$, $1.8$ and $2.2$ dB suggesting the
possibility of better schemes. With the third scheme, the corresponding numbers are $0.35$, $0.65$
and $0.85$ dB. Further reduction in these unrecovered gains could be possible with a larger
size of ${\cal C}_{\sf enh, \hsppp phase}$ or ${\cal C}_{\sf enh, \hsppp phase, \hsppp amp}$ 
(such as with $B = 3$).
However, the increase in the search space produces diminishing gains and the search complexity can
also lead to latencies associated with beam management, which can in turn translate to increased
power consumption and thermal overheads. Thus, it is of broad interest in understanding the optimal
size and structure of codebook enhancements, which could be of interest in future work.

\ignore{
\begin{figure*}[htb!]
\begin{center}
\begin{tabular}{cc}
\includegraphics[height=1.9in,width=2.7in]{fig_cbk_minus_dyncbk_0and1mmairgap_v2.eps}
&
\includegraphics[height=1.9in,width=2.7in]{fig_opt_minus_dyncbk_0and1mmairgap_v2.eps}
\\
(a) & (b)
\end{tabular}
\caption{\label{fig_cbk}
(a) Beamforming gain improvement with proposed codebook enhancement strategy over a static codebook scheme.
(b) Beamforming gain not recovered from the optimal scheme with the proposed scheme.
}
\end{center}
\end{figure*}
}

\section{Concluding Remarks}
\label{sec6} 
The scope of this work has been on understanding the implications of hand blockage at millimeter
wave frequencies. 
We first report chamber measurements performed with a commercial grade millimeter wave modem and 
a commercial grade hand phantom at $28$ GHz which captured the complete electric field information 
(array response) in
Freespace and with different blockage settings. The blockage settings correspond to the use of a hand
phantom with different spacings to the antenna module of interest and one or two fingers obstructing the
antenna elements. These studies showed that a loose hand grip-based loss estimates capture the observed
blockage losses. Further, the use of two fingers leads to more losses than one finger and the presence of
an air gap between the hand phantom and the antenna module reduces the losses.

We then quantified the performance loss between the use of a static directional
beam steering codebook of size-$4$ for the $4 \times 1$ array (typical practical deployment numbers) relative to
the optimal MRC beamforming scheme. These loss estimates showed that while the static codebook is a good
codebook for Freespace considerations, it performs relatively poorly with hand blockage. This is because
the fingers of the hand induce random phase and amplitude distortions due to multiple reflections from 
different parts of the hand, which a
static codebook cannot take advantage of. In this context, we introduced a codebook enhancement of a
quantized set of phase shifter combinations which do not carry a directional structure and simply quantize
the space of all possible phase combinations. We showed that this codebook enhancement, both theoretically 
as well as with measurement data, can lead to significant
performance improvement over the static beam steering codebook suggesting that the randomization of phases by the
presence of the hand can be de-randomized by the enhanced codebook. 

Future work on understanding the limits of such codebook enhancements taking into account latency with a 
pre-determined set of phase excitations, beam search complexity, power and thermal constraints would
be of immense practical importance. Maximum permissible exposure (MPE) constraints and the need to 
perform regulation-driven beam characterization can lead to significant complexity as the size of 
the codebook enhancement increases (e.g., $B$ of the phase shifter). Thus, decoupling the uplink and 
downlink beams at the UE side (breaking down beam correspondence~\cite{38_912spec}) and the 
tradeoffs associated with this breakdown are of broad interest. Implications of hand blockage on 
better antenna array design as well as at upper millimeter wave bands (e.g., $60$ GHz and beyond) are 
of interest. Extending these ideas to larger arrays commonly used in customer premises
equipments, integrated access and backhaul nodes, intelligent reflecting surfaces, or base-stations 
where non-blockage issues such as fading still effectively induce the same type of
amplitude and phase randomization effects~\cite{vasanth_gcom15} would also be of broader utility.

\section*{Acknowledgment}
The authors would like to thank David Henry of Qualcomm Technologies, Inc., for the design of mechanical
structures used in this study and in assistance with Fig.~1. The authors also acknowledge the feedback of 
Raghu N. Challa and Brian Banister, both of Qualcomm Technologies, Inc., on this work.

\appendix
\subsection{Proof of Theorem~\ref{thm1}}
\label{app_thm1}
First, in the high-transmit power setting, note that ${\sf S}_i$ is given as
\begin{eqnarray}
{\sf S}_i & = &
| E_ { {\sf blockage}, \hsppp i} (\theta_{ {\sf R}, \hsppp 1},
\phi_{ {\sf R}, \hsppp 1} ) |^2
\nonumber \\
& = & | E_ { {\sf free}, \hsppp i} (\theta_{ {\sf R}, \hsppp 1},
\phi_{ {\sf R}, \hsppp 1} ) |^2 \cdot A_i(\theta_{ {\sf R}, \hsppp 1},
\phi_{ {\sf R}, \hsppp 1} )^2, \hspp i = 1, \cdots, N.
\nonumber
\end{eqnarray}
Let us define the intermediate phase variable
\begin{eqnarray}
\theta_i = \angle{ E_{ {\sf free}, \hsppp i } (\theta_{ {\sf R}, \hsppp 1},
\phi_{ {\sf R}, \hsppp 1} ) } + P_i(\theta_{ {\sf R}, \hsppp 1},
\phi_{ {\sf R}, \hsppp 1} ) - \phi_{k_i}, \hspp i = 1, \cdots, N.
\nonumber
\end{eqnarray}
Since $\phi_{k_i}$ is a value from a $B$-bit phase shifter, 
we have the following bounds:
\begin{eqnarray}
|\theta_i| \leq \frac{2 \pi}{2 \cdot 2^B} \Longrightarrow
\cos \left( \frac{\pi}{2^B} \right) \leq \cos(\theta_i) \leq 1 \hspp
{\sf and} \hspp
|\sin(\theta_i)| \leq \sin \left( \frac{\pi}{2^B}
\right).
\label{eq_bounds}
\end{eqnarray}
Using the expression for ${\sf S}_i$, the achieved $\widetilde{\sf SNR}_{\sf rx}$ with 
${\cal C}_{\sf enh, \hsppp phase, \hsppp amp}$ can be seen to be
\begin{eqnarray}
\frac{1}{|\alpha_1|^2} \cdot 
\widetilde{\sf SNR}_{\sf rx}
\Big|_{ {\cal C}_{\sf enh, \hsppp phase, \hsppp amp} } & = &
\frac{
\Big( \sum_i | E_ { {\sf free}, \hsppp i} (\theta_{ {\sf R}, \hsppp 1},
\phi_{ {\sf R}, \hsppp 1} ) |^2 \cdot A_i(\theta_{ {\sf R}, \hsppp 1},
\phi_{ {\sf R}, \hsppp 1} )^2 \cos(\theta_i) \Big)^2
}
{  \sum_i  | E_ { {\sf free}, \hsppp i} (\theta_{ {\sf R}, \hsppp 1},
\phi_{ {\sf R}, \hsppp 1} ) |^2 \cdot A_i(\theta_{ {\sf R}, \hsppp 1},
\phi_{ {\sf R}, \hsppp 1} )^2 }
\nonumber \\
& & + \frac{
\Big( \sum_i | E_ { {\sf free}, \hsppp i} (\theta_{ {\sf R}, \hsppp 1},
\phi_{ {\sf R}, \hsppp 1} ) |^2 \cdot A_i(\theta_{ {\sf R}, \hsppp 1},
\phi_{ {\sf R}, \hsppp 1} )^2 \sin(\theta_i) \Big)^2 }
{  \sum_i  | E_ { {\sf free}, \hsppp i} (\theta_{ {\sf R}, \hsppp 1},
\phi_{ {\sf R}, \hsppp 1} ) |^2 \cdot A_i(\theta_{ {\sf R}, \hsppp 1},
\phi_{ {\sf R}, \hsppp 1} )^2 }.
\nonumber
\end{eqnarray}
Similarly, the achieved ${\sf SNR}_{\sf rx}$ with ${\cal C}_{\sf enh, \hsppp phase }$
can be seen to be
\begin{eqnarray}
\frac{1}{|\alpha_1|^2} \cdot  {\sf SNR}_{\sf rx}
\Big|_{ {\cal C}_{\sf enh, \hsppp phase } } & = &
 \frac{1}{N} \cdot \left(
\sum_i
| E_ { {\sf free}, \hsppp i} (\theta_{ {\sf R}, \hsppp 1},
\phi_{ {\sf R}, \hsppp 1} ) | \cdot A_i(\theta_{ {\sf R}, \hsppp 1},
\phi_{ {\sf R}, \hsppp 1} ) \cdot \cos(\theta_i) \right)^2
\nonumber \\
& & +
\frac{1}{N} \cdot \left(
\sum_i
| E_ { {\sf free}, \hsppp i} (\theta_{ {\sf R}, \hsppp 1},
\phi_{ {\sf R}, \hsppp 1} ) | \cdot A_i(\theta_{ {\sf R}, \hsppp 1},
\phi_{ {\sf R}, \hsppp 1} ) \cdot \sin(\theta_i) \right)^2. 
\nonumber
\end{eqnarray}
Using~(\ref{eq_bounds}), we have the following inequalities:
\begin{align}
&
\frac{
\Big( \sum_i | E_ { {\sf free}, \hsppp i} (\theta_{ {\sf R}, \hsppp 1},
\phi_{ {\sf R}, \hsppp 1} ) |^2 \cdot A_i(\theta_{ {\sf R}, \hsppp 1},
\phi_{ {\sf R}, \hsppp 1} )^2 \cos(\theta_i) \Big)^2
}
{  \sum_i  | E_ { {\sf free}, \hsppp i} (\theta_{ {\sf R}, \hsppp 1},
\phi_{ {\sf R}, \hsppp 1} ) |^2 \cdot A_i(\theta_{ {\sf R}, \hsppp 1},
\phi_{ {\sf R}, \hsppp 1} )^2 }
\nonumber \\
& {\hspace{0.2in}}
\geq \cos^2 \left( \frac{\pi}{2^B} \right) \cdot
\sum_i | E_ { {\sf free}, \hsppp i} (\theta_{ {\sf R}, \hsppp 1},
\phi_{ {\sf R}, \hsppp 1} ) |^2 \cdot A_i(\theta_{ {\sf R}, \hsppp 1},
\phi_{ {\sf R}, \hsppp 1} )^2 
{\hspace{0.05in}} {\sf and} 
\nonumber \\
&
\frac{
\Big( \sum_i | E_ { {\sf free}, \hsppp i} (\theta_{ {\sf R}, \hsppp 1},
\phi_{ {\sf R}, \hsppp 1} ) |^2 \cdot A_i(\theta_{ {\sf R}, \hsppp 1},
\phi_{ {\sf R}, \hsppp 1} )^2 \sin(\theta_i) \Big)^2
}
{  \sum_i  | E_ { {\sf free}, \hsppp i} (\theta_{ {\sf R}, \hsppp 1},
\phi_{ {\sf R}, \hsppp 1} ) |^2 \cdot A_i(\theta_{ {\sf R}, \hsppp 1},
\phi_{ {\sf R}, \hsppp 1} )^2 } \geq 0.
\nonumber
\end{align}
We also have
\begin{eqnarray}
\left(\sum_i
| E_ { {\sf free}, \hsppp i} (\theta_{ {\sf R}, \hsppp 1},
\phi_{ {\sf R}, \hsppp 1} ) | \cdot A_i(\theta_{ {\sf R}, \hsppp 1},
\phi_{ {\sf R}, \hsppp 1} ) \cdot \cos(\theta_i) \right)^2
\leq
\left( \sum_i
| E_ { {\sf free}, \hsppp i} (\theta_{ {\sf R}, \hsppp 1},
\phi_{ {\sf R}, \hsppp 1} ) | \cdot A_i(\theta_{ {\sf R}, \hsppp 1},
\phi_{ {\sf R}, \hsppp 1} ) \right)^2   \nonumber \\
\left|
\sum_i
| E_ { {\sf free}, \hsppp i} (\theta_{ {\sf R}, \hsppp 1},
\phi_{ {\sf R}, \hsppp 1} ) | \cdot A_i(\theta_{ {\sf R}, \hsppp 1},
\phi_{ {\sf R}, \hsppp 1} ) \sin(\theta_i)
\right| \leq \sin\left( \frac{\pi}{2^B} \right)  \cdot
\sum_i
| E_ { {\sf free}, \hsppp i} (\theta_{ {\sf R}, \hsppp 1},
\phi_{ {\sf R}, \hsppp 1} ) | \cdot A_i(\theta_{ {\sf R}, \hsppp 1},
\phi_{ {\sf R}, \hsppp 1} ).
\nonumber
\end{eqnarray}
Putting these inequalities together, we have 
\begin{eqnarray} 
\widetilde{\Delta} {\sf SNR}_{\sf rx} 
& \geq & 
\sum_i | E_ { {\sf free}, \hsppp i} (\theta_{ {\sf R}, \hsppp 1},
\phi_{ {\sf R}, \hsppp 1} ) |^2 \cdot A_i(\theta_{ {\sf R}, \hsppp 1},
\phi_{ {\sf R}, \hsppp 1} )^2 \cdot \cos^2 \left( \frac{ \pi}{2^B} \right) 
\nonumber \\ 
& & 
-\frac{1}{N} \left( \sum_i
| E_ { {\sf free}, \hsppp i} (\theta_{ {\sf R}, \hsppp 1},
\phi_{ {\sf R}, \hsppp 1} ) | \cdot A_i(\theta_{ {\sf R}, \hsppp 1},
\phi_{ {\sf R}, \hsppp 1} ) \right)^2 
\nonumber \\ 
& & - \frac{1}{N} \sin^2 \left( \frac{ \pi}{2^B} \right) 
\cdot \left( \sum_i
| E_ { {\sf free}, \hsppp i} (\theta_{ {\sf R}, \hsppp 1},
\phi_{ {\sf R}, \hsppp 1} ) | \cdot A_i(\theta_{ {\sf R}, \hsppp 1},
\phi_{ {\sf R}, \hsppp 1} ) \right)^2. 
\nonumber 
\end{eqnarray} 
The statement in~(\ref{eq_thm1}) is straightforward upon simplification of the above 
expression.
\endproof

\bibliographystyle{IEEEbib}
\bibliography{newrefsx2new}


\end{document}